# Nanoscale spin rectifiers for harvesting ambient radiofrequency energy

Raghav Sharma[1,2], Tung Ngo[1], Eleonora Raimondo[3], Anna Giordano[4], Junta Igarashi[5], Butsurin Jinnai[6], Shishun Zhao[1], Jiayu Lei[1], Yong-Xin Guo[1], Giovanni Finocchio[3], Shunsuke Fukami[5-8], Hideo Ohno[5-8] and Hyunsoo Yang[1*]

[1]*Department of Electrical and Computer Engineering, National University of Singapore, 117576, Singapore*

[2]*Department of Electrical Engineering, Indian Institute of Technology Ropar, Rupnagar 140001, India*

[3]*Department of Mathematical and Computer Sciences, Physical Sciences and Earth Sciences, University of Messina, I-98166 Messina, Italy*

[4]*Department of Engineering, University of Messina, I-98166 Messina, Italy*

[5]*Laboratory for Nanoelectronics and Spintronics, Research Institute of Electrical Communication, Tohoku University, 2-1-1 Katahira, Aoba, Sendai 980-8577, Japan*

[6]*Advanced Institute for Materials Research (WPI-AIMR), Tohoku University, Sendai 980-8577, Japan*

[7]*Center for Science and Innovation in Spintronics, Tohoku University, 2-1-1 Katahira, Aoba, Sendai 980-8577, Japan*

[8]*Center for Innovative Integrated Electronic Systems, Tohoku University, 468-1 Aramaki Aza Aoba, Sendai 980-0845, Japan*
*e-mail: eleyang@nus.edu.sg

**Radiofrequency harvesting using ambient wireless energy could be used to reduce the carbon footprint of electronic devices. However, ambient radiofrequency energy is weak (less than −20 dBm), and the performance of state-of-the-art radiofrequency rectifiers is restricted by thermodynamic limits and high-frequency parasitic impedance. Nanoscale spin rectifiers based on magnetic tunnel junctions have recently demonstrated high sensitivity, but suffer from a low a.c.-to-d.c. conversion efficiency (less than 1%). Here, we report a sensitive spin rectifier rectenna that can harvest ambient radiofrequency signals between −62 and −20 dBm. We also develop an on-chip co-planar waveguide-based spin rectifier**

**array with a large zero-bias sensitivity (around 34,500 mV mW⁻¹) and high efficiency (7.81%). The self-parametric excitation driven by voltage-controlled magnetic anisotropy is a key mechanism that contributes to the performance of the spin-rectifier-array. We show that these spin rectifiers can be used to wirelessly power a sensor at a radiofrequency power of −27 dBm.**

Wireless sensor networks (WSNs) play a critical role in applications such as health and environmental monitoring, and the Internet of Things (IoT). Sensors installed in difficult to access locations – which are required, for example, in monitoring air quality, temperature, and moisture – should ideally be battery-free. Large sensor and IoT device networks require numerous radiofrequency (rf) sources to exchange data. A considerable amount of the ambient rf energy from these sources remains unused and could thus be harvested and converted to dc power[1,2] to power electronic devices and sensors[3-8] (Fig. 1a). Compared to other energy harvesting sources (such as solar, vibration, thermal, and wind), rf energy harvesting offers all-day availability, easy accessibility, and can be integrated with small WSNs[9,10].

A rf energy harvesting module (EHM) consists of a receiver antenna, a rectifier, a power management module, and a load that uses the harvested dc power (Fig. 1b). The EHM should work at an ambient available rf power level ($P_{rf}$) of below −20 dBm (10 µW), as per the IEEE wireless LAN standards for the 2.4 GHz band (which is the most abundant waste rf source in the ambient)[11]. The rectifier is an important part of the EHM, and defines the overall efficiency ($P_{dc}/P_{rf}$, where $P_{dc}$ is the harvested dc power). State-of-the-art rf rectifiers are based on Schottky diodes, transistors, and complementary metal–oxide–semiconductor (CMOS) technology[12-14]. The Schottky diodes (either on rigid substrates such as silicon and III–V compound semiconductors[15,16] or flexible substrate based on two-dimensional $MoS_2$[14]) and tunnel diodes[17], are the most efficient[13,14,18,19] rectifiers for the electromagnetic energy harvesting. Efficiencies can reach 40–70% [5,6,14,20-25], but in the low-power regime ($P_{rf}$ less than −20 dBm), the thermodynamic limit creates a challenge[26,27]. Schottky diodes (with an antenna) have achieved an efficiency of 5–19% (2.45 GHz) at $P_{rf}$ between −30 and −20 dBm[5,24,28-33], but practical implementation with an EHM operating at $P_{rf}$ less than −20 dBm remains a challenge[3,5,6,14,34]. A commercial 2.45 GHz EHM has been developed by Powercast, but only works at $P_{rf}$ larger than -10 dBm[34].

Spin rectifiers (SRs) based on magnetic tunnel junctions (MTJs) could overcome the low-power limitations of charge-based rectifiers and eliminate antenna and matching circuits SRs can convert a rf to a d.c. signal due to the spin diode effect[35,36], where an applied rf current exerts a torque on the local spins of the free magnetic layer of the SRs, resulting in a precession motion and resistance oscillation at the same frequency as the rf signal. The resistance oscillation with a rf current produces a rectified voltage ($V_r$) across the top and bottom contacts of the SR (Fig. 1c). SRs are nano-sized, sensitive[27], CMOS compatible[37], battery-free[38] and frequency-tunable[36], making them a potential candidate for next-generation rectifiers. Sensitivities in the range of $1.2 \times 10^4$–$10^6$ mV mW$^{-1}$ have been reported in dc-biased SRs[35] using a perpendicular free layer[27], canted free-layer[38,39], vortex-expulsion[40,41] and a spin bolometer[42]. Furthermore, SR-based EHMs can operate at $P_{rf}$ around 0 dBm at 2.45 GHz[7], but with an efficiency of less than 1%, which is impractical for harvesting at ambient condition. Further advances in performance are thus needed before SRs can be used in applications such as detectors and sensors[4,7,37,43]. In particular, the narrow rectification band of less than 2 GHz [4,38,39,42] needs to be improved and the efficiency needs to be increased[4].

In this Article, we report on two SR prototypes: a 2.45 GHz SR rectenna for low-power energy harvesting at $P_{rf}$ between −62 dBm and −20 dBm, and a SR array integrated with an on-chip co-planar waveguide for broadband rectification with a zero-bias sensitivity (*S*) of around 34,500 mV mW$^{-1}$ and an efficiency of 7.81%. Micromagnetic simulations and microwave emissions data show that the approach relies on self-parametric excitations driven by voltage-controlled magnetic anisotropy (VCMA), which reduces the threshold for excitation of a broadband rectification response and leads to enhanced sensitivity in the array arrangement of SRs. Our SRs can be integrated into EHMs, and their signal-to-noise ratio (SNR) and noise equivalent power (NEP) performances are suitable for low microwave power detector applications in noisy environments.

**Device engineering for low power operation**

We use the uniformly magnetized SRs for GHz-range rectification. Our SRs consist of a CoFeB free layer (FL) and reference layer (RL) separated by an MgO spacer layer (Fig. 1c). The free and spacer layer thickness and SRs dimension are varied to control the interfacial anisotropy and tunneling magnetoresistance (TMR), in order to have the proper values of zero-bias resistance, resonance frequency, and equilibrium direction of the magnetization. We find that the SRs with the FL and spacer thickness of 1.9 nm and 1 nm, respectively, and dimensions of 40 × 100 nm$^2$ to 80 × 200 nm$^2$ demonstrate the best rectification results (Supplementary Notes 1-3).

We study the detailed properties with the zero-bias and zero magnetic field conditions. Figure 1d shows the rectification response (rectified voltage vs. input microwave frequency) of 40 × 100 nm$^2$ and 80 × 200 nm$^2$ devices at $P_{rf} = -30$ dBm. We select the SRs with the frequency close to WiFi (2.4 GHz), 4G (2.3-2.6 GHz) and 5G (3.5 GHz) bands. We observe that the rectified signal in the 40 × 100 nm$^2$ to 80 × 200 nm$^2$ SRs is a combination of symmetric and anti-symmetric Lorentzian[44].

We find the maximum sensitivity of ~3800 mV mW$^{-1}$ and ~2200 mV mW$^{-1}$ at $P_{rf}$ = −55 dBm and −40 dBm, respectively, for the 80 × 200 nm$^2$ and 40 × 100 nm$^2$ SRs, which is higher than the zero-bias sensitivity of previous reports[7,27,38,44] and a state-of-the-art zero-bias Schottky diode[45]. The better sensitivity and rectification voltage in the 80 × 200 nm$^2$ SRs are due to a high zero-field TMR, $\Delta R_{zb}$ = $(R_{zb} - R_p)/R_p$,[27,38] (Supplementary Fig. 2) and larger VCMA (Supplementary Note 4), where $R_{zb}$ is the zero-field resistance and $R_P$ is the parallel configuration resistance. The 80 × 200 nm$^2$ SR shows VCMA > 50 fJ/Vm and $\Delta R_{zb}$ > 30%. A high zero-field TMR ($\Delta R_{zb}$) more than 30% is useful for acquiring a large zero-field and zero-bias rectified signal. Moreover, a significant VCMA > 50 fJ/Vm improves the sensitivity due to the nonlinear contribution[44] (Supplementary Note 4), where the observed VCMA is stronger than the reported value of 30-50 fJ/Vm in MTJ-based SRs[38,44,46]. Recent improvements of TMR to 631%[47] and the VCMA to 320 fJ Vm$^{-1}$ [48] may enhance the rectified voltage of SRs. A relatively low resistance-area (RA) product is another key requirement, as a large RA significantly reduces the rf current flow through the SRs. Furthermore, to produce a significant zero-field and zero-bias rectified signal, canted anisotropy with a $\Delta R_{zb}$ is required. Therefore, canted SRs with a high zero-field TMR, low RA, and high VCMA coefficient are useful for energy harvesting applications. The sensitivity decreases at $P_{rf}$ > −20 dBm (Fig. 1e) due to the decrease of $\Delta R_{zb}$ and saturation of the rectified voltage ($V_r$)[7,27,38] (Supplementary Note 5).

**SR-rectenna for ultra-sensitive rf energy harvesting**

While injecting an rf signal using a probe station, a rf power loss of 35–60% due to impedance mismatching is observed at 2.4 GHz (Supplementary Note 6). To improve sensitivity, a high-gain impedance matched antenna ($R_x$) is designed and attached to the 80 × 200 nm$^2$ SR (Fig. 2a), to utilize them as a SR-rectenna. Figure 2b shows the antenna design, where the transmission line dimensions are varied to match the SR impedance at 2.45 GHz (Supplementary Note 7). The rectified voltage and sensitivity in Fig. 2c,d shows significant enhancement due to an impedance-matched SR-rectenna, especially at ultra-low power $P_{rf} < -50$ dBm. The sensitivity of probe station-based excitation in Fig. 1e and SR-rectenna-based wireless excitation in Fig. 2d shows similar results (shown in red), indicating that the SR-rectenna is well integrated. Using the matched SR-rectenna, the rectified voltage is $V_r = 6.27$ µV at $P_{rf} = -62$ dBm, corresponding to $S$ ~10,000 mV mW$^{-1}$, which is larger than the $S$ of 450-1850 mV mW$^{-1}$ reported in other SRs[7,27,38] and the unmatched SR (3800 mV mW$^{-1}$) in Fig. 1e. A high sensitivity more than 1000 mV mW$^{-1}$ is observed for a wide range of $P_{rf}$ between −62 and −25 dBm.

**Role of magnetic anisotropy in designing SR-array**

The matched SR-rectenna shows a significantly improved rf sensitivity. However, the voltage is insufficient for the EHM. Hence, it is necessary to connect SRs in an array for enhancing the output voltage. Compared to Schottky diode cascading (2 to 4 stages) for voltage multiplication, the SRs can be easily connected by the small wirebonds[7,37], making them compact. Moreover, a potential on-chip integration of SRs with the CMOS technology is expected to have more benefits from our SR-array demonstration. Here, we discuss the influence of single SR behavior on the array design using $40 \times 100$ nm$^2$ and $80 \times 200$ nm$^2$ devices.

First, we consider the experimental rectification response of the individual SR. In contrast to the resonant behavior at $P_{rf} \leq -30$ dBm (Fig. 1d), the $40 \times 100$ nm$^2$ SR shows a broadband response (0.1–3.5 GHz) at $P_{rf} \geq -25$ dBm (Fig. 3a), whereas the $80 \times 200$ nm$^2$ SR shows a resonant behavior at both $P_{rf} \geq -25$ dBm (Fig. 3b) and $P_{rf} \leq -30$ dBm (Fig. 1d). In addition, due to tunability with the magnetic field, a large bandwidth of 0.1–6 GHz using a single SR is observed (Supplementary Fig. 8a). This bandwidth is twice of previous reports of SRs[4,49].

To study the scalability of the SR technology, we connect 10 SRs ($40 \times 100$ nm$^2$ and $80 \times 200$ nm$^2$) in series without any external antenna to excite the SR-array directly by wireless rf energy, which is the key challenge for the development of ultra-compact EHMs. To save on-chip area for making ultra-compact EHMs, a small co-planar waveguide ($240 \times 500$ μm$^2$) is attached to individual SRs to couple the rf power. The $40 \times 100$ nm$^2$ SR-array shows a broadband response compatible with the individual SRs response at $P_{rf} > -25$ dBm (Fig. 3a). In contrast to the single SR broadband behaviour at a high magnetic field (Supplementary Fig. 8a), a SR-array due to the enhanced self-parametric effect achieves a wide range broadband behaviour without any magnetic field (Fig. 3a), making it more appropriate for on-chip applications. Compared to the $40 \times 100$ nm$^2$ SR-array, the $80 \times 200$ nm$^2$ SR-array shows a transition from resonant to broadband with a narrower detection bandwidth (Fig. 3b) but higher sensitivity. Hence, the $80 \times 200$ nm$^2$ SR-array can be used for a band-pass filter-rectifier. Overall, the peak rectified voltage is scaled by approximately $N$ times upto 4-5 devices, where $N$ is the number of the SRs connected in series. After that, the output starts to deviate from the $N$ scaling and begins to saturate (Supplementary Fig. 9). In general, more number of SRs connected in series show a broadband rectification response, and hence increase the integrated rectification voltage.

We perform the micromagnetic simulations to reproduce the transition from resonant to broadband in a $40 \times 100$ nm$^2$ single SR. Fig. 3c summarizes a phase diagram of the rectification curves as a function of the frequency and amplitude of the current density ($J_{ac}$) where the transition from resonant to broadband detection can be observed. Fig. 3d shows the amplitude of the $x$-component of the magnetization, $\Delta m_x$ (parallel to the polarizer) as a function of the amplitude of $J_{ac}$. Examples of simulated rectification curves are displayed in Fig. 3e. It can be noted that at a low input power (e.g. $J_{ac} = 0.1$ MA/cm$^2$) in Fig. 3e, the detection curves are resonant with a small precession angle. As the power increases (e.g. $J_{ac}$ = 10 MA/cm$^2$), a large amplitude magnetization precession $\Delta m_x$ leads to the broadband response. This rectification mechanism is observed in small SRs (40 × 100 nm$^2$), where the canted equilibrium angle is 50-70°. On the other hand, for the 80 × 200 nm$^2$ SR, where the canted equilibrium angle is close to out-of-plane (79-95°), the transition from resonant to broadband is not observed[7,38].

In order to understand the potential origin of the transition from resonant to broadband response in the 40 × 100 nm$^2$ SR-array, we have performed additional experimental studies. The microwave emissions of a single and two series connected 40 × 100 nm$^2$ SRs driven by a rf current $i_{ac} = I_{ac} \sin(2\pi f t)$ at a frequency of $f$ are recorded in the spectrum analyzer (Fig. 3f,g). The microwave emissions of a single SR exhibit a weak second harmonics (2$f$ component) at 6 GHz (Fig. 3f), whereas the second harmonics peak power is enhanced by one order of magnitude in the two SRs connected in series (Fig. 3g). The decrease in the threshold power required for the onset of 2$f$ peak i.e., the self-parametric effect and the enhancement of 2$f$ peak power in the series-connected SRs is shown in Fig. 3h. The presence of the second harmonics can be understood considering that the magnetoresistance oscillates at the same frequency of the input rf current $R(t) = \Delta R_S \sin(2\pi f t + \varphi_R)$, where $\varphi_R$ is a phase shift. The voltage across the SR is given by

$$V_r = \frac{1}{2}\Delta R_S I_{ac} \cos(\varphi_R)(1 - \cos(2\pi f t + \varphi_R)) + \frac{1}{2}\Delta R_S I_{ac} \sin(\varphi_R) \sin(2\pi f t + \varphi_R). \quad (1)$$

As already demonstrated, these devices have large VCMA, when connected in series, the voltage at 2$f$ originated in one SR drives a parametric excitation in the other SR enhancing the magnetization precession angle. This mechanism recalls the self-parametric excitation[50,51] and it can explain the large voltage (large sensitivity) observed in Fig. 3b. To support this claim, we have performed a systematic micromagnetic study of the magnetization oscillation as a function of the VCMA amplitude considering an rf current at $f$ and a VCMA applied at 2$f$. Supplementary Fig.

10a shows an example of these calculation, comparing the response at VCMA field of 0 mT and 10 mT. The calculation results confirm that the amplitude of magnetization precession increases with increasing VCMA, supporting the idea of self-parametric excitation. The VCMA-based self-parametric effect opens a path for design of next generation of high-performance SRs. In addition, we expect that the microwave magnetic field of the incident electromagnetic wave can couple directly to the free layer magnetization enhancing the amplitude of the magnetization precession.

**SR-array based broadband low-powered EHM**

Due to the high sensitivity and rectified voltage in the 80 × 200 $nm^2$ SR-array (10 SRs in series), we integrate them in an EHM as shown in Fig. 4a, in an ambient environment without an extra external antenna. The SR-array shows the rectification voltage ($V_r$) of 20 mV and 11 mV using the 2.45 GHz and 3.5 GHz rf sources at $P_{rf}$ = −25 dBm, respectively (Fig. 4b). The voltage threshold of the boost converter is 20 mV, which SR-array achieves at −22 dBm at 2.45 GHz.

The SR-array voltage ($V_r$) steps up from 20–50 mV to 1.6–4 V by the boost converter. The temperature sensor turns on at $V_{step}$ ~1.2 V, which is achieved at −27 dBm and −22 dBm using the dual sources (2.45 GHz and 3.5 GHz) and a single 2.45 GHz source, respectively, for the EHM demonstration. The EHM takes a total 15–30 seconds for powering the sensor initially. The response time of SR-array to accumulate ~20 mV takes only a few seconds due to the low capacitance (pF to fF), which is another major advantage of SRs. However, to avoid the fast charging-discharging effect of SRs due to a low capacitance, we use an external capacitor of 0.01 F (capacity 3.3 V) to hold a stable rectified voltage. The SR-array voltage is $V_r$ ~28 mV at $P_{rf}$ = −25 dBm using those two sources, which is converted to $V_{step}$ ~3.7 V by the boost converter. The estimated charging time of an external capacitor to reach 20 mV at $P_{rf}$ = −25 dBm is ~23-24 seconds, which is close to the experimentally observed charging time of around 20-25 seconds for the EHM to turn on the sensor. Here, the differential resistance of the SR-array is measured as ~2 k$\Omega$. EHM can work steadily for one hour and the $V_{step}$ can be stored in the capacitor (Supplementary Note 11). The $V_{step}$ is then connected to a low-current driven handheld multimeter, showing a negligible discharging or leakage effect after turning off the wireless source (Fig. 4c). The 1.2 V temperature sensor and 1.6 V LED can remain ON for ~50 s and 30 s, respectively, after turning off the wireless source, useful for discontinuous rf environment. Hence, SR-array based EHM can be used for broadband operation to power various electronic devices.

**Comparison with existing rf energy harvesting technology**

In 2014, Hemour *et al.*[13] predicted that an optimized SR could outperform state-of-the-art rectifiers in the rf power range of −70 to −10 dBm. We compare the SR-rectenna and SR-array with the two commercially available low-power Schottky diodes (HSMS-2860 and SMS-7630) under the same ambient condition in Fig. 5a,b. We measure the zero-bias differential resistance ($R_{zbr}$) of SR, SR-array, HSMS 2860, and SMS 7630 when irradiated by 2.45 GHz at $P_{rf}$ = −25 dBm and find the values of 0.27, 2, 5, and 5.1 k$\Omega$, respectively. Here, $P_{rf}$ is the power emitted from the signal generator, without including any antenna gain and losses at the device interface due to impedance mismatching to determine the actual front-end efficiency (see method). Our results show that SR-rectenna works reliably (SNR > 20 dB) for −62 dBm < $P_{rf}$ < −20 dBm, enabling SR-rectenna sensitive for the weak ambient condition (Supplementary Note 12). Furthermore, we measure the noise equivalent power (NEP) of a single SR, which is defined as the ratio of $V_{noise}$/sensitivity, where $V_{noise}$ is the noise voltage. A low NEP of ~2 × 10$^{-12}$ W Hz$^{0.5}$ is measured at $P_{rf}$ = 0.1 μW (−40 dBm) without a dc bias ($I_{dc}$ = 0 mA), which is slightly smaller than the NEP measured in previous reports of SRs with a dc bias[38] (Supplementary Note 12). The observed high SNR and low NEP imply that SRs are suitable for low-power microwave detector applications.

The SR-array shows a low SNR at $P_{rf} < -50$ dBm in comparison to SR-rectenna due to an increased parasitic capacitance (Supplementary Fig. 12a). However, in the rf power range of −50 dBm to −20 dBm, an SR-array outperforms due to a high efficiency ($\eta$ ~7.81% at −30 dBm) and sensitivity (~34,500 mV mW$^{-1}$ at −50 dBm) as shown in Fig. 5a,b. The efficiency is 2-3 orders of magnitude higher than the previous SR (1.2 GHz)[4] at $P_{rf} < -20$ dBm (Supplementary Fig. 13).

Although the performance of state-of-the-art rf rectifiers based on Si, GaAs or MoS$_2$, can reach an efficiency of 40–80% at $P_{rf} > -10$ dBm[6,14,20-25], their performance at $P_{rf} < -20$ dBm is limited (Supplementary Fig. 14) for the EHM demonstrations as summarized in Table 1. Only few works show a high-efficiency of 5–19%[5,24,28-33] at −30 dBm < $P_{rf} \leq -20$ dBm after including the antenna efficiency, whereas our work shows an SR-array efficiency without antenna in Fig. 5a.

Furthermore, the SR-array is connected in an equivalent electrical area of ~1 mm$^2$, including the co-planar waveguide, which couples the rf power to the SR devices. Packaged Schottky diodes (> 10 mm$^2$) with a high-efficiency antenna captures an overall area of at least few cm$^2$, impractical for on-chip scaling. Even with the miniaturized antenna, the size is reduced only to ~200 mm$^2$ for low-power applications[5]. Our demonstration shows that the SR-array could be the most-sensitive and compact solution for ambient energy harvesting technology (Table 1).

## Conclusions

We have reported a SR rectenna with a large rf sensitivity of around 10,000 mV mW$^{-1}$ at −62 dBm, which can harvest rf energy in weak and noisy ambient environments. In the case of a single SR, intrinsic device properties such as the perpendicular anisotropy, device geometry, and dipolar field from the polarizer layer play a crucial role in defining the energy landscape of the nanomagnet, and the excitation of a large angle magnetization precession at low input power. The sensitivity is linked to the dynamical response of the MTJ and depends on the zero-field TMR and the VCMA coefficient, which enhances the zero-bias rectified voltage driven by the spin-polarized current. Furthermore, an external matching setup can minimize input power losses and enhance the sensitivity of the SR-rectenna. In the case of the SR array, we observed a VCMA-driven self-parametric effect, which results in an enhancement of the sensitivity and detection bandwidth without incorporating any external antenna or matching setup. We showed that our SR-array-based EHM can power a commercial sensor at a low rf power of −27 dBm without the use of an extra antenna. Our SRs are compact, not prone to suffering from parasitic effects, easy to integrate, scalable, and efficient in ambient conditions.

## Methods

**Device structure.** The SR layered structure is sapphire substrate/Ta (5)/Ru (10)/Ta (5)/Pt$_{38}$Mn$_{62}$ (15)/Co (2.4)/Ru (0.88)/Co$_{18.75}$Fe$_{56.25}$B$_{25}$ (2.4)/MgO ($t_{MgO}$)/Co$_{18.75}$Fe$_{56.25}$B$_{25}$ ($t_{CoFeB}$)/Ta (5)/Ru (5) (thicknesses in nm). The MgO and other layers are deposited using the rf and dc magnetron sputtering, respectively, at room temperature and base pressure $< 1 \times 10^{-6}$ Pa with the Ar gas flow. The thickness of topmost CoFeB free layer ($t_{CoFeB}$) and spacer MgO layer ($t_{MgO}$) are varied from 1.7–1.9 nm and 0.9–1.2 nm, respectively, to tune the canted equilibrium direction (along the $z$-direction from the $x$ direction in Fig. 1c) as described in Supplementary Notes 1-3. The bottom CoFeB (2.4 nm) is the reference layer (RL), whose magnetization direction is pinned along the -$x$ direction due to Co/Ru/CoFeB synthetic antiferromagnetic (SAF) coupling, which is exchange biased by an antiferromagnet PtMn.

**Device engineering.** The SRs were first optimized by varying $t_{CoFeB}$ and $t_{MgO}$. The SRs with $t_{CoFeB}$ = 1.9 nm and $t_{MgO}$ = 1 nm, show the best performance. For further tuning the SRs performance, the stack is processed into elliptical SRs with the dimensions of 40 × 100 nm² to 160 × 400 nm² using e-beam lithography and Ar ion beam milling by keeping the aspect ratio of 2.5. The 40 × 100 nm² to 80 × 200 nm² shows the best zero-bias rectification results due to canted equilibrium magnetization. The bigger devices (100 × 250 nm² to 160 × 400 nm²) show low rectification performance. Here, we use the 40 × 100 nm² and 80 × 200 nm² for discussing the scalability and diverse functionalities of SR. The best optimized 80 × 200 nm² SR devices show the first-order anisotropy ($\mu_0 H_{k,eff}$) of 0.25–0.3 T, TMR of 53–82%, RA of 4–6 Ω μm², frequency of 2.1–3.6 GHz, and resistance < 500 Ω.

**Microwave measurement for the optimization.** For the optimization of SRs, we use the probe-station for the rf excitation and rectification measurements. The SRs rectification responses are measured using the spin torque ferromagnetic resonance (ST-FMR) technique. A rf signal modulated at a low frequency of 213 Hz is applied using a signal generator and the resulting rectified dc voltage is measured by a lock-in amplifier with a time constant of 100 milliseconds and a low pass filter slope of 24 dB/octave. The rf loss of 35-60% on the linear scale in the frequency range from 2.4 to 2.5 GHz is measured due to the impedance mismatch of 50 Ω terminated probes and SRs using the reflection coefficient ($S_{11}$) measured on the linear scale in the vector network analyser. We have included the rf loss in the sensitivity extraction for the probe station-based measurements in Fig. 1e and Supplementary Fig. 3d, where the sensitivity is calculated as $S = V_r/P$. Here, $P = P_{rf} \times (1-|S_{11}|^2)$, is the power reaching the SR-devices after including the losses and $P_{rf}$ is the actual power fed from the signal generator (rf source) to the wires in the probe station-based measurements.

**Rectenna Design.** The rectenna is designed by attaching an antenna to the SR. For the antenna design shown in Fig. 2b, we first measure the $S_{11}$ of the SRs using VNA for the impedance calculation (Supplementary Fig. 7a). The patch antenna is fabricated on the RO4003C substrate.

The dimension of the transmission line is optimized to match the SR impedance at 2.45 GHz. The insertion loss of the matched antenna is −17.5 dB at 2.45 GHz and can cover the entire range of the WiFi band (2.4–2.5 GHz) with $S_{11} < -10$ dB (Supplementary Fig. 7b). A high gain of 7.2 dBi is measured in the far-field region (Supplementary Fig. 7c). The SR-rectenna is tested successfully for a distance of 10–15 cm between the transmitter and receiver antenna. Another high-gain 50 Ω patch antenna is designed for the comparison. The coupling between the transmitting and receiving antenna used in the SR-rectenna is determined by the VNA measurement as shown in Supplementary Fig. 7d,e. The sensitivity for the rectenna measurement is $S = V_r/P_{rf}$, where $P_{rf}$ is determined by the $S_{21}$ parameter of the transmitting and receiving antennas (Supplementary Fig. 7d,e).

**Spectrum analyzer measurement:** In Fig. 3f,g, we have used the spectrum analyzer measurement. The details of the setup are given in ref. 7. In the spectrum analyser, the SRs oscillation signal and reflected signal due to the input rf signal ($P_{rf}$) coexist. Even though the SRs oscillation signal is separated from the injected rf input signal through a directional coupler, the separation is imperfect. Since the linewidth of SRs (few MHz) and reflected signal (few Hz) from the injected rf input are very different, we can separate the injected input signal from the SR signal as a background subtraction. As the resolution of the spectrum analyzer is a few kHz (much larger than the reflection of the injected rf signal linewidth), the injected rf signal appears as a single data point, which could be easily subtracted by a proper Lorentzian fitting of the measured signal.

**Multi-band rectification and EHM demonstration.** The broadband rectification in Fig. 3a,b is performed using a high-gain (6–7 dBi) broadband (0.8–18 GHz) horn antenna. For the multi-band EHM demonstration, we used a resonant 2.4–2.5 GHz patch antenna (gain ~7.2 dBi) and 3.5–3.6 GHz whip antenna (gain ~3 dBi). The antennas are fed by the signal generator. The SR-array voltage ($V_r$) is stored in the capacitor before amplified by the boost converter to 1.5–4 V. For the stability analysis in Fig. 4c, the step-up voltage ($V_{step}$ ~3.7 V) is first stored and then supplied to the multimeter, LED, and temperature sensor. The EHM demonstration is performed at varying distances of 2.5–5 cm. For the EHM demonstration at –27 dBm, the antenna is placed at ~2.5 cm from the SR-array chip. The co-planar waveguide design, on chip devices, and setup component images are shown in Supplementary Fig. 15.

**Comparison with Schottky diode.** For the comparison, the HSMS-2860 and SMS-7630 Schottky diodes are connected with the 50 Ω antenna, whereas a single SR is attached with an impedance matched antenna. The efficiency is calculated by $\eta = P_{dc}/P_{rf}$, where $P_{dc} = V_r^2/R_{zbr}$. Here, $R_{zbr}$ is the resistance of the SR-devices or Schottky diodes monitored during the measurement. The sensitivity is calculated as $S = V_r/P_{rf}$. Here, for a fair evaluation of all the systems together, the $P_{rf}$ is the actual power emitted from the rf source (signal generator), without correcting for the transmitting antenna gain and losses at the device interface or due to attenuation to estimate a front-end efficiency. Please note that the sensitivity calculation includes the losses only for probe station measurements used for optimization (Fig. 1e), while no losses are taken into account for the rectenna and energy harvesting measurements, underestimating the sensitivity. In the independent measurements using the VNA ($S_{21}$), spectrum analyser connected with the receiving antenna, or rf power meter to measure the emitted wireless power from the transmitting antenna, we found that the power available at 2.5 cm differs from the power fed by the rf source ($P_{rf}$) only by ± 0.5-3 dBm, which shows the high directivity of the transmitting antenna with minimal gain and attenuation in the near field region. Hence, in the near-field measurements, we find that the received power at 2.5 cm is very close to the actual power fed by the signal generator ($P_{rf}$) and no correction due to any antenna gain or loss at the device interface is considered for the sensitivity estimation, independent of using SR-rectenna, SR-array or Schottky diodes for the wireless energy harvesting. Since the losses at the device interface are ignored, the derived sensitivity denotes the lower bound of the device performance. The evaluated performances of the HSMS-2860 and SMS-7630 Schottky diodes are well matched with their data sheets. The SR-array is tested without an antenna, where the small ground-signal-ground co-planar waveguide (240 µm × 500 µm) of Cr (5 nm)/Au (100 nm) formed by photolithography and lift-off on the SR, couples the rf signals. The small wirebonds (< 1 mm), used to connect the MTJs, may couple a small amount of rf power to the MTJs. However, these wirebonds are mainly used as a dc or low frequency transmission path for the rectified signal.

**Micromagnetic simulations.** Micromagnetic simulations have been performed by solving numerically the Landau-Lifshitz-Gilbert-Slonczewski equation[52]:

$$\frac{d\bm{m}}{dt} = \frac{\gamma M_S}{(1+\alpha_G^2)}\left[-(\bm{m}\times\bm{h}_{eff}) - \alpha_G(\bm{m}\times\bm{m}\times\bm{h}_{eff}) + \sigma J(\bm{m}\times\bm{m}\times\bm{p})\right] \quad (2)$$

where $\alpha_G$ is the Gilbert damping, $\bm{m} = \bm{M}/M_S$ is the normalized magnetization vector, $\gamma$ is the gyromagnetic ratio and $M_S$ is the saturation magnetization of the MTJ free layer (FL). The effective magnetic field, $\bm{h}_{eff}$, includes the demagnetizing field, and the interfacial uniaxial perpendicular anisotropy. The spin-transfer torque (STT) term is proportional to the pre-factor $\sigma = \frac{g|\mu_B|P}{|e|\gamma_0 M_{S_1}^2 d_z}$, where $g$ is the gyromagnetic splitting factor, $\mu_B$ is the Bohr magneton, $e$ is the electron charge, $d_z$ is the thickness of the FL, and $P$ is the spin polarization. The ac current density flowing into the device is given by $J = J_{ac}\sin(2\pi f t + \varphi_J)$. The VCMA is included in the effective field as a changing anisotropy field. The amplitude of the VCMA field can be computed as $h_{VCMA} = \frac{2\xi V_b}{\mu_0 M_S^2 t_z t_{ox}}$, where $\xi$ is the linear VCMA coefficient, $M_s$ is the saturation magnetization, $t_z$ is the free layer thickness, and $t_{ox}$ is the oxide thickness. For a single device, the VCMA field $h_{VCMA}$ can be larger than 2 mT ($M_s$ = 800 kA/m, $t_z$ = 1.9 nm, $t_{ox}$ = 1 nm, microwave current density of $1 \times 10^6$ A/cm², $R_{AP}$ = 1 kΩ, and $\xi$ = 50 fJ/(mV)), where $R_{AP}$ is the resistance in anti-parallel configuration. However, in order to simulate the effect of self-parameter excitation at twice the input microwave frequency, which enhances the microwave emissions power as shown in Fig. 3h, the effect of the VCMA field is analyzed for a higher, more reasonable value of 10 mT in Supplementary Fig. 10a.

We have simulated an elliptical 40 nm × 100 nm × 1 nm hybrid device, where the perpendicular anisotropy is larger than the demagnetizing magnetic field and hence the equilibrium direction of the magnetization **m** is tilted from the out-of-plane direction as observed in the experiment. The polarizer magnetization **p** is in-plane along the -x direction (Fig. 1c). The main micromagnetic parameters are $M_S = 800$ kA/m, perpendicular uniaxial anisotropy constant $K_u = 0.39$ MJ/m³, $\alpha_G = 0.02$, $g = 2$, and $P = 0.7$, and the dipolar field originated by the polarizer is added as a constant $H_{DC} = -5$ mT. In simulations the rectification curves are computed for the amplitude of the ac current density raging from $J_{ac} = 0.1$ to 10 MA/cm² and a frequency scan from $f = 0.1$ to 15 GHz. In the simulations presented in Supplementary Fig. 10a, the frequency of the VCMA is two times of that of the applied ac current density. For matching the frequency of the experimental data of rectification curves for the 40 × 100 nm² single SR ($f \sim 3.5$ GHz) in Fig. 3e, the perpendicular uniaxial anisotropy constant is tuned to $K_u = 0.419$ MJ/m³ (Supplementary Fig. 10b).

**Table 1** | Comparison of state-of-the-art 2.45 GHz rf rectifiers for the low-power operation and EHM demonstration.

| | Area (rectifier + antenna) | Low power Efficiency | Antenna | Power for EHM demo | Ref |
|---|---|---|---|---|---|
| MoS$_2$ flexible rectifier | ~few mm$^2$ | ~2% at −20 dBm | On-chip antenna | +3 dBm | [14] |
| HSMS-2852 | 214.1 mm$^2$ | ~20% at −20 dBm (including antenna efficiency) | Rectenna | +10 dBm | [5] |
| HSMS-2852 | Few cm$^2$ | ~7−8% at −25 dBm | Rectenna | +36 dBm | [6] |
| Powercast commercial module | Full board | 5 % at −10 dBm | Separate antenna | > −10 dBm | [34] |
| 65 nm CMOS technology | CMOS Chip | − | 1.27 cm$^2$ antenna | +15.5 dBm | [53] |
| SR-array | 1.2 mm$^2$ | 7.8% at −27 dBm | On-chip co-planar waveguide (no external antenna) | −27 dBm | This work |

References

1 Shaikh, F. K. & Zeadally, S. Energy harvesting in wireless sensor networks: A comprehensive review. *Renew. Sust. Energ. Rev.* **55**, 1041-1054 (2016).
2 Shaikh, F. K., Zeadally, S. & Exposito, E. Enabling technologies for green internet of things. *IEEE Syst. J.* **11**, 983-994 (2015).
3 Bito, J., Kim, S., Tentzeris, M. & Nikolaou, S. Ambient energy harvesting from 2-way talk-radio signals for "smart" meter and display applications. *2014 IEEE Antennas and Propagation Society International Symposium (APSURSI)*, 1353-1354 (2014).
4 Fang, B. *et al.* Experimental demonstration of spintronic broadband microwave detectors and their capability for powering nanodevices. *Phys. Rev. Appl.* **11**, 014022 (2019).
5 Koohestani, M., Tissier, J. & Latrach, M. A miniaturized printed rectenna for wireless RF energy harvesting around 2.45 GHz. *AEU - Int. J. Electron.* **127**, 153478 (2020).
6 Olgun, U., Chen, C.-C. & Volakis, J. L. Wireless power harvesting with planar rectennas for 2.45 GHz RFIDs. *2010 URSI International symposium on electromagnetic theory*, 329-331 (2010).
7 Sharma, R. *et al.* Electrically connected spin-torque oscillators array for 2.4 GHz WiFi band transmission and energy harvesting. *Nat. Commun.* **12**, 2924 (2021).
8 Vallem, V., Sargolzaeiaval, Y., Ozturk, M., Lai, Y. C. & Dickey, M. D. Energy harvesting and storage with soft and stretchable materials. *Adv. Mater.* **33**, 2004832 (2021).
9 Iqbal, N., Masood, M., Nasir, A. A. & Qureshi, K. K. Review of contemporary energy harvesting techniques and their feasibility in wireless geophones. *Int. J. Energy Res.* **46**, 5703-5730 (2022).
10 Tran, L.-G., Cha, H.-K. & Park, W.-T. RF power harvesting: a review on designing methodologies and applications. *Micro Nano Syst. Lett.* **5**, 1-16 (2017).
11 Committee, I. C. S. L. M. S. IEEE Standard for Information technology-Telecommunications and information exchange between systems-Local and metropolitan area networks-Specific requirements Part 11: Wireless LAN Medium Access Control (MAC) and Physical Layer (PHY) Specifications. *IEEE Std 802.11* (2007).
12 Hemour, S. & Wu, K. Radio-frequency rectifier for electromagnetic energy harvesting: Development path and future outlook. *Proc. IEEE* **102**, 1667-1691 (2014).
13 Hemour, S. *et al.* Towards low-power high-efficiency RF and microwave energy harvesting. *IEEE Trans. Microw. Theory Tech.* **62**, 965-976 (2014).
14 Zhang, X. *et al.* Two-dimensional MoS2-enabled flexible rectenna for Wi-Fi-band wireless energy harvesting. *Nature* **566**, 368-372 (2019).
15 Strohm, K. M., Buechler, J. & Kasper, E. SIMMWIC rectennas on high-resistivity silicon and CMOS compatibility. *IEEE Trans. Microw. Theory Tech.* **46**, 669-676 (1998).
16 Suh, Y.-H. & Chang, K. A high-efficiency dual-frequency rectenna for 2.45-and 5.8-GHz wireless power transmission. *IEEE Trans. Microw. Theory Tech.* **50**, 1784-1789 (2002).
17 Lorenz, C. H. P. *et al.* Breaking the efficiency barrier for ambient microwave power harvesting with heterojunction backward tunnel diodes. *IEEE Trans. Microw. Theory Techn.* **63**, 4544-4555 (2015).
18 Cansiz, M., Altinel, D. & Kurt, G. K. Efficiency in RF energy harvesting systems: A comprehensive review. *Energy* **174**, 292-309 (2019).

19  Assogba, O., Mbodji, A. K. & Diallo, A. K. in *2020 IEEE International Conf on Natural and Engineering Sciences for Sahel's Sustainable Development-Impact of Big Data Application on Society and Environment (IBASE-BF).*  1-10 (IEEE).
20  Chen, Y.-S. & Chiu, C.-W. Maximum achievable power conversion efficiency obtained through an optimized rectenna structure for RF energy harvesting. *IEEE Trans. Antennas Propag.* **65**, 2305-2317 (2017).
21  Kim, J. & Jeong, J. Design of high efficiency rectifier at 2.45 GHz using parasitic canceling circuit. *Microw. Opt. Technol. Lett.* **55**, 608-611 (2013).
22  Mbombolo, S. E. F. & Park, C. W. in *2011 IEEE MTT-S International Microwave Workshop Series on Innovative Wireless Power Transmission: Technologies, Systems, and Applications.*  23-26 (IEEE).
23  Olgun, U., Chen, C.-C. & Volakis, J. L. Investigation of rectenna array configurations for enhanced RF power harvesting. *IEEE Antennas Wirel. Propag. Lett.* **10**, 262-265 (2011).
24  Shi, Y. *et al.* An efficient fractal rectenna for RF energy harvest at 2.45 GHz ISM band. *Int. J. RF Microw.* **28**, e21424 (2018).
25  Wang, D. & Negra, R. Design of a rectifier for 2.45 GHz wireless power transmission. *PRIME 2012; 8th Conference on Ph. D. Research in Microelectronics & Electronics*, 1-4 (2012).
26  Valenta, C. R. Fundamental limitations for Schottky diode RF energy harvesting. *2015 IEEE International Conference on RFID Technology and Applications (RFID-TA)*, 188-193 (2015).
27  Miwa, S. *et al.* Highly sensitive nanoscale spin-torque diode. *Nat. Mater.* **13**, 50-56 (2014).
28  Valenta, C. R. & Durgin, G. D. Harvesting wireless power: Survey of energy-harvester conversion efficiency in far-field, wireless power transfer systems. *IEEE Microw. Mag.* **15**, 108-120 (2014).
29  Vera, G. A., Georgiadis, A., Collado, A. & Via, S. Design of a 2.45 GHz rectenna for electromagnetic (EM) energy scavenging. *2010 IEEE Radio and Wireless Symposium (RWS)*, 61-64 (2010).
30  Assimonis, S. D., Fusco, V., Georgiadis, A. & Samaras, T. Efficient and sensitive electrically small rectenna for ultra-low power RF energy harvesting. *Sci. Rep.* **8**, 15038 (2018).
31  Adami, S.-E. *et al.* A flexible 2.45-GHz power harvesting wristband with net system output from− 24.3 dBm of RF power. *IEEE Trans. Microw. Theory Tech.* **66**, 380-395 (2017).
32  Song, C. *et al.* A high-efficiency broadband rectenna for ambient wireless energy harvesting. *IEEE Trans. Antennas Propag.* **63**, 3486-3495 (2015).
33  Shen, S., Zhang, Y., Chiu, C.-Y. & Murch, R. A triple-band high-gain multibeam ambient RF energy harvesting system utilizing hybrid combining. *IEEE Trans. Ind. Electron.* **67**, 9215-9226 (2019).
34  *P21XXPowerharvester Chipset Reference Design Evaluation Board*, <www.powercastco.com/documentation/p21xxcsr-evb-datasheet/> (2018).
35  Finocchio, G. *et al.* Perspectives on spintronic diodes. *Appl. Phys. Lett.* **118**, 160502 (2021).
36  Tulapurkar, A. *et al.* Spin-torque diode effect in magnetic tunnel junctions. *Nature* **438**, 339-342 (2005).
37  Romera, M. *et al.* Vowel recognition with four coupled spin-torque nano-oscillators. *Nature* **563**, 230-234 (2018).

38   Fang, B. *et al.* Giant spin-torque diode sensitivity in the absence of bias magnetic field. *Nat. Commun.* **7**, 11259 (2016).
39   Zhang, L. *et al.* Ultrahigh detection sensitivity exceeding $10^5$ V/W in spin-torque diode. *Appl. Phys. Lett.* **113**, 102401 (2018).
40   Jenkins, A. *et al.* Spin-torque resonant expulsion of the vortex core for an efficient radiofrequency detection scheme. *Nat. Nanotechnol.* **11**, 360-364 (2016).
41   Tsunegi, S. *et al.* Achievement of high diode sensitivity via spin torque-induced resonant expulsion in vortex magnetic tunnel junction. *Appl. Phys. Express* **11**, 053001 (2018).
42   Goto, M. *et al.* Uncooled sub-GHz spin bolometer driven by auto-oscillation. *Nat. commun.* **12**, 536 (2021).
43   Leroux, N. *et al.* Radio-frequency multiply-and-accumulate operations with spintronic synapses. *Phys. Rev. Appl.* **15**, 034067 (2021).
44   Zhu, J. *et al.* Voltage-induced ferromagnetic resonance in magnetic tunnel junctions. *Phys. Rev. Lett.* **108**, 197203 (2012).
45   *Zero-Bias_Schottky_Diode_Detectors_100kHz-50Ghz.pdf*, <http://www.herotek.com/datasheets> (2022).
46   Nozaki, T. *et al.* Electric-field-induced ferromagnetic resonance excitation in an ultrathin ferromagnetic metal layer. *Nat. Phys.* **8**, 491-496 (2012).
47   Scheike, T., Wen, Z., Sukegawa, H. & Mitani, S. 631% room temperature tunnel magnetoresistance with large oscillation effect in CoFe/MgO/CoFe(001) junctions. *Appl. Physics. Lett.* **122**, doi:10.1063/5.0145873 (2023).
48   Nozaki, T. *et al.* Highly efficient voltage control of spin and enhanced interfacial perpendicular magnetic anisotropy in iridium-doped Fe/MgO magnetic tunnel junctions. *NPG Asia Mater.* **9**, e451-e451, doi:10.1038/am.2017.204 (2017).
49   Zhang, L. *et al.* Enhanced broad-band radio frequency detection in nanoscale magnetic tunnel junction by interface engineering. *ACS Appl. Mater. Interfaces* **11**, 29382-29387 (2019).
50   Nitzan, S. H. *et al.* Self-induced parametric amplification arising from nonlinear elastic coupling in a micromechanical resonating disk gyroscope. *Sci. Rep.* **5**, 9036 (2015).
51   Rhodes Jr, J. Parametric self-excitation of a belt into transverse vibration. *J. Appl. Mech.* **37**, 1055-1060 (1970).
52   Tomasello, R. *et al.* Low-frequency nonresonant rectification in spin diodes. *Phys. Rev. Appl.* **14**, 024043 (2020).
53   Sadagopan, K. R., Kang, J., Ramadass, Y. & Natarajan, A. A 960 pW co-integrated-antenna wireless energy harvester for WiFi backchannel wireless powering. *2018 IEEE International Solid-State Circuits Conference-(ISSCC)*, 136-138 (2018).

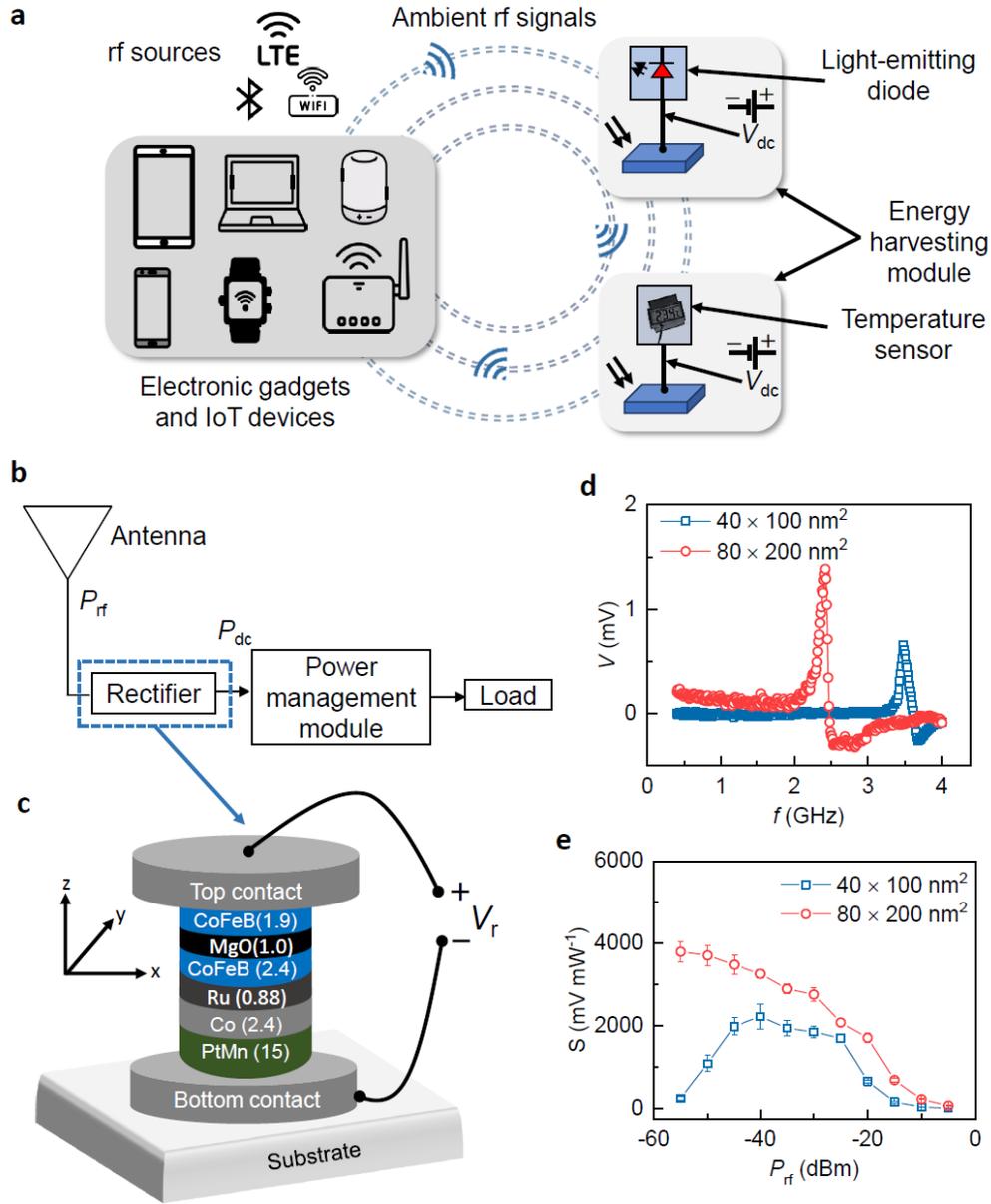

**Fig 1| The rf energy harvesting using spin-rectifiers. a**, Illustration of the energy harvesting modules powering up small sensors and electronic components such as a light-emitting diode and temperature sensor by converting ambient rf energy (shown by dashed circles) to a useful dc voltage ($V_{dc}$). The ambient rf energy is shared by various electronic gadgets and internet of things (IoT) devices, whereas WiFi routers and transmitters for the Bluetooth and Long-Term Evolution (LTE) represent the rf sources. **b**, Prototype model of the energy harvesting module, where an ambient rf power ($P_{rf}$) is converted to a dc power ($P_{dc}$) that can be used as a usable electric power by the load. **c**, Layered structure of spin-rectifier with top and bottom contacts. **d**, Zero-bias and zero magnetic field rectification ($V$) as a function of frequency ($f$) of $40 \times 100$ nm$^2$ and $80 \times 200$ nm$^2$ devices at $P_{rf} = -30$ dBm. **e**, Sensitivity ($S$) at 3.5 GHz and 2.45 GHz from $40 \times 100$ nm$^2$ and $80 \times 200$ nm$^2$ devices, respectively as a function of the rf power ($P_{rf}$). The error bar in the $S$ values shows the standard deviation of the peak rectified voltage from five measurements.

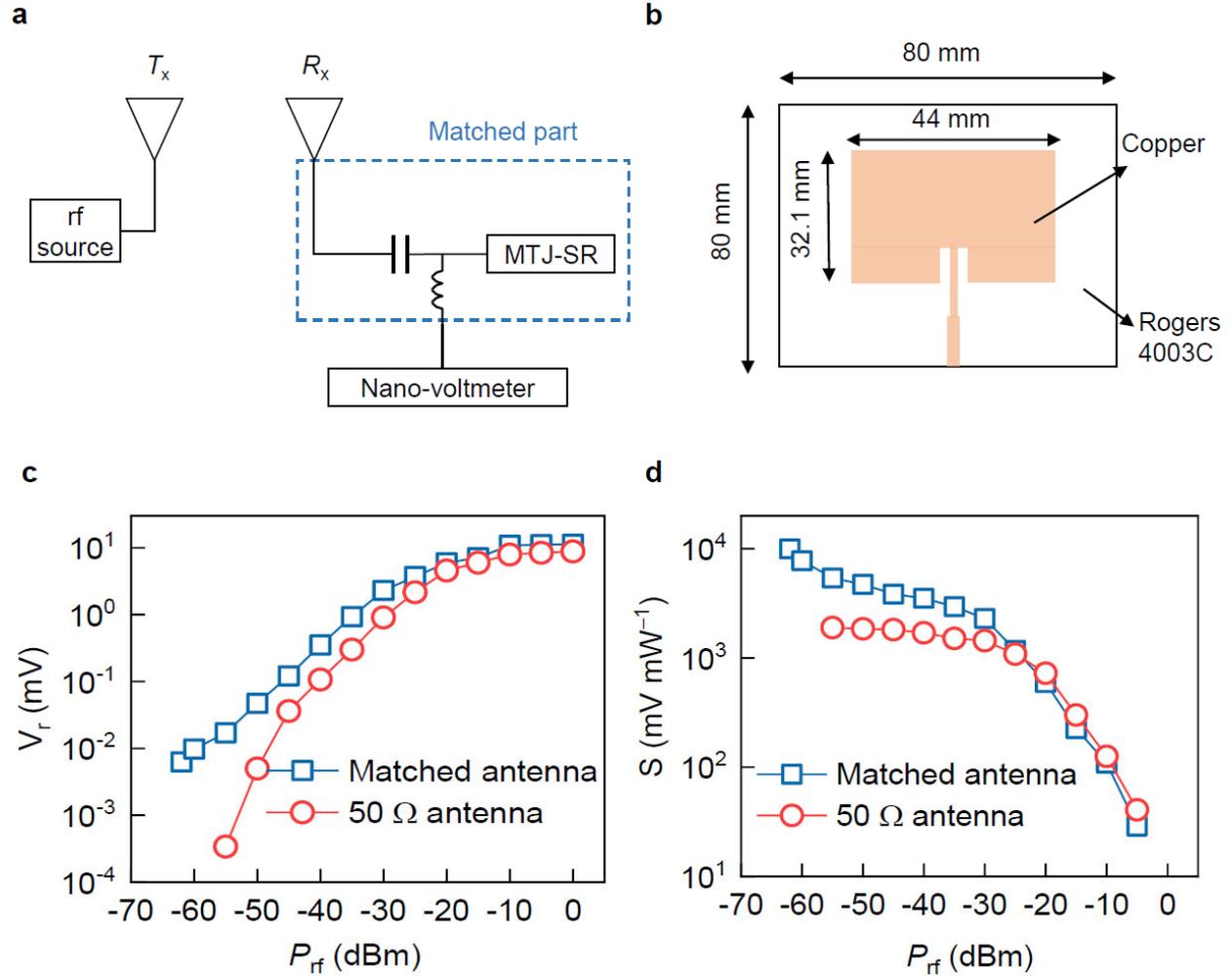

**Fig. 2| Performance of SR-rectenna. a,** Schematic of integration of the matched antenna with the SR. $T_x$ is the 50 Ω transmitting antenna and $R_x$ is the impedance matched receiving antenna. Another 50 Ω receiving antenna is also designed for comparison. **b,** Patch antenna designed to match the impedance of the matched part in **a**. **c,d,** Rectified voltage ($V_r$) and sensitivity (S) comparison of 50 Ω and impedance matched receiving antenna attached to the 80 × 200 nm² SR with varying the rf power ($P_{rf}$) at 2.45 GHz. To ensure the stable measurement of $V_r$ and corresponding S, the $V_r$ is measured after a waiting time of 5 seconds to reach the peak voltage and then averaged over 30 seconds. The standard deviation in $V_r$ and S values in rectenna measurement is less than 2% over time.

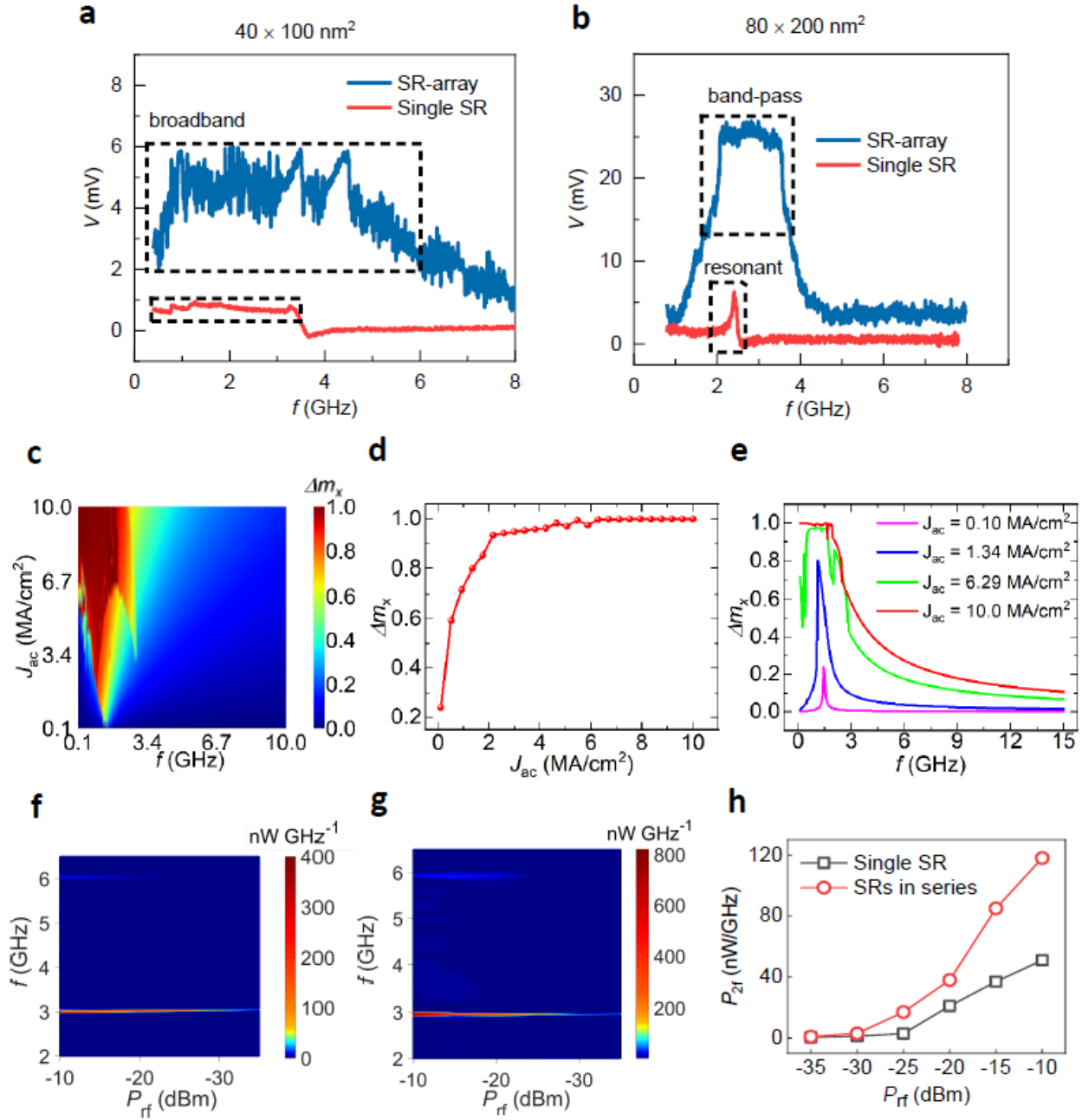

**Fig. 3| Tuning of broadband and resonant recfication for designing SR-array. a**, Experimental results of broadband rectification voltage ($V$) versus frequency ($f$) response from a 40 × 100 nm² single SR and an array of ten SRs connected in series at rf power ($P_{rf}$) of −20 dBm. The dashed rectangular boxes approximately show the region of broadband rectification, where the rectifition voltage is significant for the energy harvesting from 100 MHz to 3.5 GHz and 100 MHz to 6 GHz for the single SR and SR-array, respectively. **b**, Resonant and band-pass behavior of a 80 × 200 nm² single SR and an array of ten SRs connected in series, respectively, at $P_{rf}$ = −20 dBm. The rectangular boxes approximately shows the regions of resonant behaviour (a combination of symmetric and anti-symmetric Lorentzian) and band-pass behaviour for the single SR and SR-array, respectively. **c**, Simulated phase diagram of the precession amplitude of the *x*-component

($\Delta m_x$) of the magnetization (component originating the TMR) as a function of the microwave frequency ($f$) and current density amplitude ($J_{ac}$). **d**, Simulated results of the maximum precession $\Delta m_x$ as a function of $J_{ac}$ for the 40 × 100 nm² single SR. **e**, Simulated $\Delta m_x$ as a function of $f$ at various $J_{ac}$ with perpendicular uniaxial anisotropy constant, $K_u$ = 0.419 MJ/m³, revealing the transition from resonant to broadband behaviour. **f,g**, Comparison of experimentally observed spectrum emitted from the single (**f**) and two series connected SRs (**g**) excited by a rf signal ($I_{dc}$ = 0 mA). **h**, Extracted 2$f$ peak power ($P_{2f}$) shows an enhancement in the case of SRs in series. The $P_{2f}$ is calculated from the time-averaged data using the trace averaging function in the spectrum analyser over 5 sweeps.

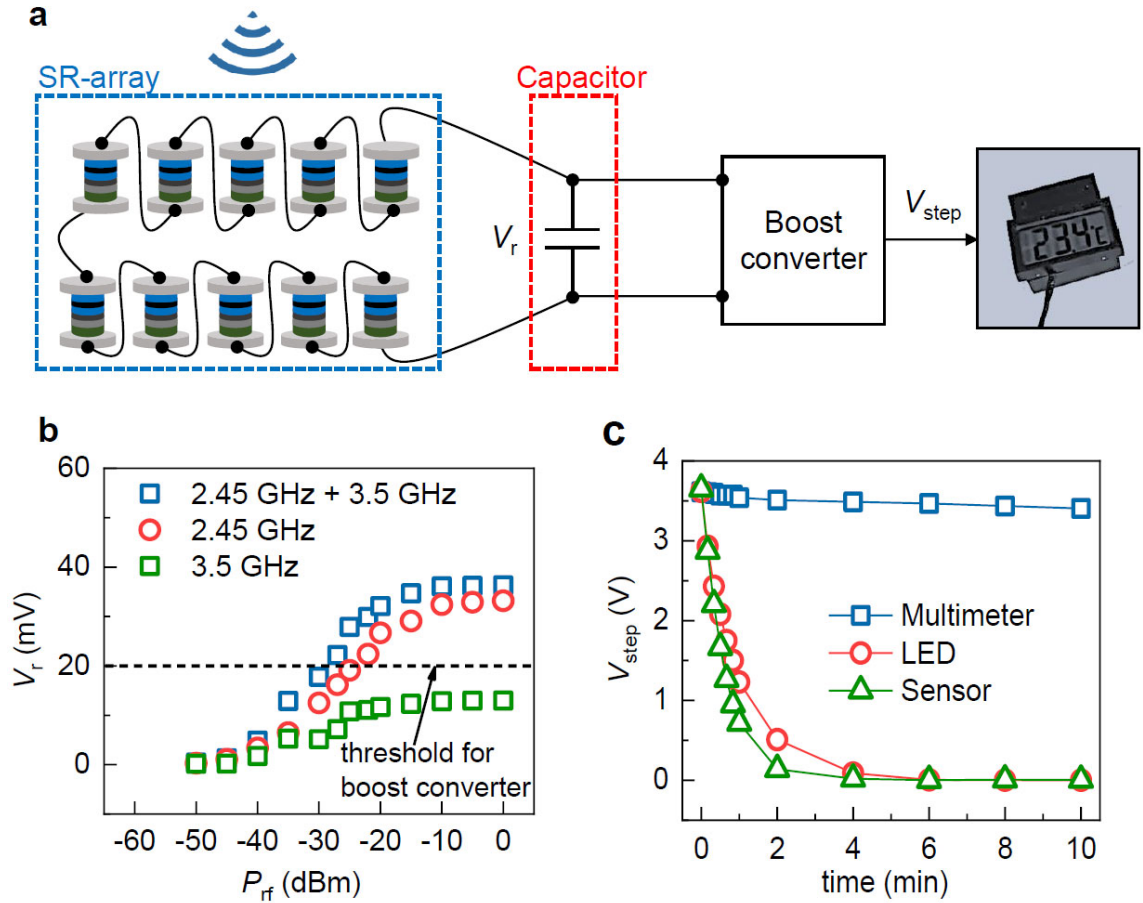

**Fig. 4| Demonstration of broadband low-power SR based energy harvesting module. a**, A schematic of the energy harvesting module, where the harvested voltage ($V_r$) from the SR-array is first stored in the capacitor, which is stepped up to a high-voltage ($V_{step}$ ~1.6–4V) using the boost converter to power the temperature sensor. The ON state of temperature sensor corresponds to $V_{step}$ > 1.2 V. **b**, Rectified voltage ($V_r$) from the SR-array at various rf powers ($P_{rf}$) using 2.45 GHz and 3.5 GHz antenna. The rectified voltage at each $P_{rf}$ is recorded after 5 seconds waiting time to reach the peak voltage and then averaged over next 30 seconds to ensure the stable charging of the capacitor via the SR-array generated rectified voltage. The standard deviation in the measured $V_r$ is less than 4% over time, as shown in Supplementary Fig. 11a. The dashed line at $V_r$ ~ 20 mV is the threshold voltage for the ON state of boost converter. **c**, Discharging of stored $V_{step}$ when connected to different loads in the EHM after turning off the rf power source. Here, the SR-array is first irradiated by two rf sources, 2.45 GHz and 3.5 GHz antenna, simultaneously at $P_{rf}$ = −25 dBm for at least 30 seconds to reach the peak voltage, and then the rf sources are turned off for measuring the discharging $V_{step}$. The $V_{step}$ represents a standard deviation of less than 1% over time, as shown in Supplementary Fig. 11b. **c**, Discharging of stored $V_{step}$ when connected to different loads in the EHM after turning off the rf power source. Here, the SR-array is first irradiated by two rf sources, 2.45 GHz and 3.5 GHz antenna, simultaneously at $P_{rf}$ = −25 dBm for at least 30 seconds to reach the peak voltage, and then the rf sources are turned off for measuring the discharging $V_{step}$. The $V_{step}$ represents a standard deviation of less than 1% over time, as shown in Supplementary Fig. 11b.

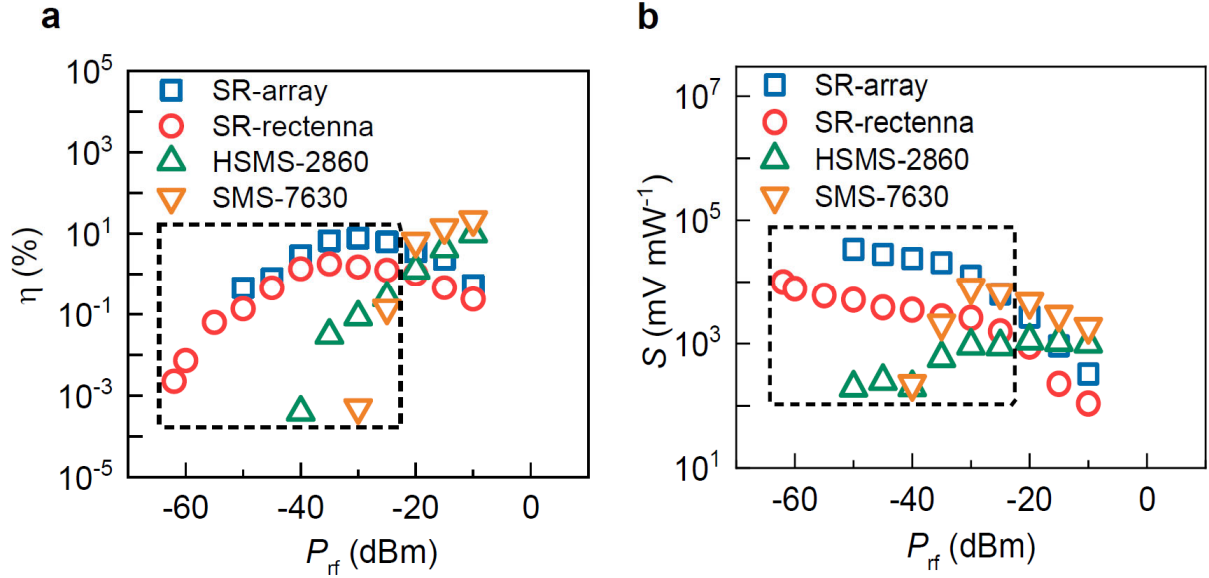

**Fig. 5| Comparison of the rectification performance of Schottky diodes, SR-array and SR-rectenna. a,b**, The conversion efficiency, $\eta$ (**a**) and sensitivity, $S$ (**b**) at 2.45 GHz with varying the rf power ($P_{rf}$). The dashed rectangular boxs show the region of rf power, where the performance of SR-array and SR-rectenna surpass that of Schottky diodes. The Schottky diodes and a single SR are integrated with an antenna to form a rectenna, whereas the SR-array directly harvests the rf energy from the wireless source. The performance of both the Schottky diodes are well matched with literature (Supplementary Fig. 14). All rectifiers are under the same ambient condition.

# Supplementary Information

**Supplementary Note 1. Optimization of interfacial anisotropy**

The optimization of MgO (tunnel barrier)/CoFeB (free layer) interface in magnetic tunnel junction (MTJ) based spin-rectifiers (SRs) are required to manipulate the interfacial perpendicular anisotropy (IPA) and to achieve a high rectification sensitivity without the magnetic field and dc bias. For this purpose, the CoFeB ($t_{CoFeB}$) and MgO ($t_{MgO}$) thicknesses are varied to tune the IPA. We have observed a canted equilibrium direction due to moderate IPA at $t_{CoFeB}$ = 1.9 nm and $t_{MgO}$ = 1 nm, a high tunneling magnetoresistance (Fig. S1a), a resonance frequency close to ~2 GHz (Fig. S1b) and a high sensitivity (Fig. S1c) using an elliptical dimension of 80 × 200 nm$^2$.

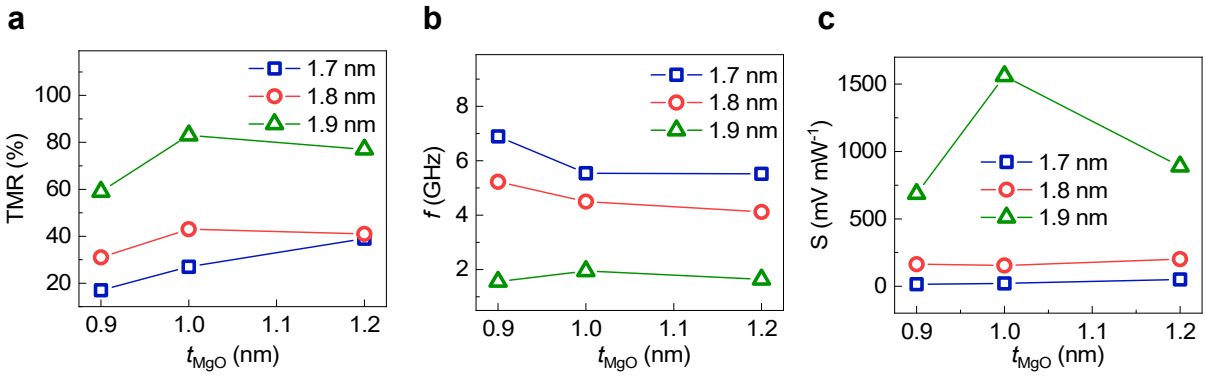

**Fig. S1| Optimization of 80 × 200 nm$^2$ SRs for the zero-bias and zero magnetic field operation with varying the thicknesses of the free layer ($t_{CoFeB}$) and spacer layer ($t_{MgO}$). a**, Tunnelling tunneling magnetoresistance (TMR) versus $t_{MgO}$. **b**, Resonance frequency ($f$) corresponding to the maximum rectification value at $P_{rf}$ = −30 dBm with varying $t_{MgO}$. **c**, The rf sensitivity ($S = V_r/P_{rf}$) at the resonance frequency at $P_{rf}$ = −30 dBm as a function of $t_{MgO}$, where $V_r$ is the maximum rectification value corresponding to the resonant frequency. The thickness of the free layer ($t_{CoFeB}$) displayed inside the plots is varied from 1.7 to 1.9 nm.

**Supplementary Note 2. Optimization of the canted equilibrium direction of SRs**

Beside the thickness optimization, we have optimized the junction dimensions of SRs for further controlling the equilibrium direction of the out-of-plane tilted magnetization at zero field. The canted equilibrium angle primarily arises from the competition of in-plane shape anisotropy and out-of-plane interfacial anisotropy. The canted equilibrium angle ($\theta_{tilt}$) between the magnetization of the in-plane reference layer and that of the tilted free layer can be estimated by[1,2]

$$R(\theta_{tilt}) = \frac{R_P + R_{AP}}{2} + \frac{R_P - R_{AP}}{2}\cos(\theta_{tilt}) \quad (S1)$$

Here, $R(\theta_{tilt})$, $R_P$, and $R_{AP}$ are the tilted, parallel, and anti-parallel configuration resistances, respectively. We have seen a significant canted equilibrium angle of 48° and 80° for the small-dimension devices: 40 × 100 nm² to 80 × 200 nm² (Fig. S2a,b), respectively. For bigger dimension devices (>100 × 250 nm²), the out-of-plane anisotropy is significantly low, favoring in-plane easy-axis anisotropy. Also, the zero-field TMR ($\Delta R_{zb} = (R_{zb} - R_p)/R_p$) reaches 30–40% in some of the 80 × 200 nm² SR devices as shown in Fig. S2b, resulting into a high rectified voltage. Here, $R_{zb}$ is the zero-field resistance.

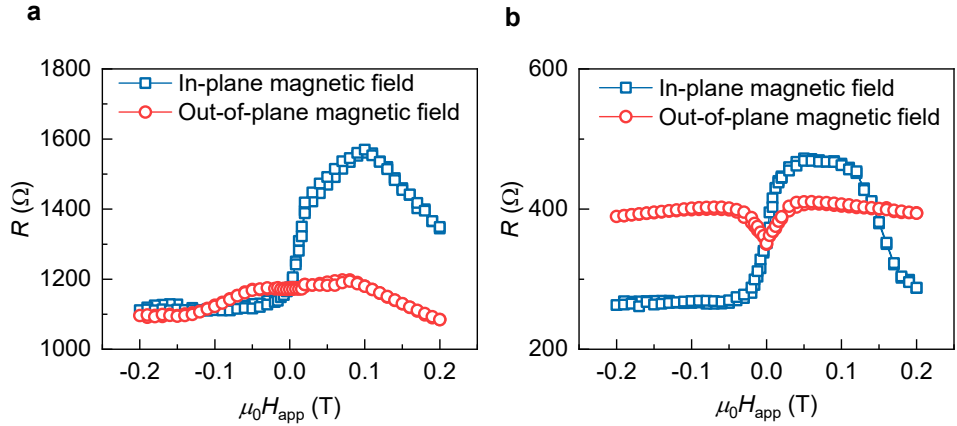

**Fig S2| Resistance ($R$) versus applied magnetic field ($\mu_0H_{app}$).** The canted equilibrium anisotropy of SRs due to competition between in-plane shape anisotropy and out-of-plane interfacial anisotropy is inferred from samples with a size of 40 × 100 nm² (**a**) and 80 × 200 nm² (**b**).

**Supplementary Note 3. Optimization of SR size for the best rf sensitivity**

The SRs with dimensions 80 × 200 nm² using $t_{CoFeB}$ = 1.9 nm and $t_{MgO}$ = 1 nm showed the best TMR of ~75% and a resistance-area product of ~5 Ω µm² (Fig. S3a). Other characteristics are a canted equilibrium angle of 79–95° (Fig. S3b), a ferromagnetic resonance (FMR) frequency close to 2.4 GHz and a maximum sensitivity of ~1000 mV mW$^{-1}$ at zero bias and zero magnetic field (Fig. S3c,d).

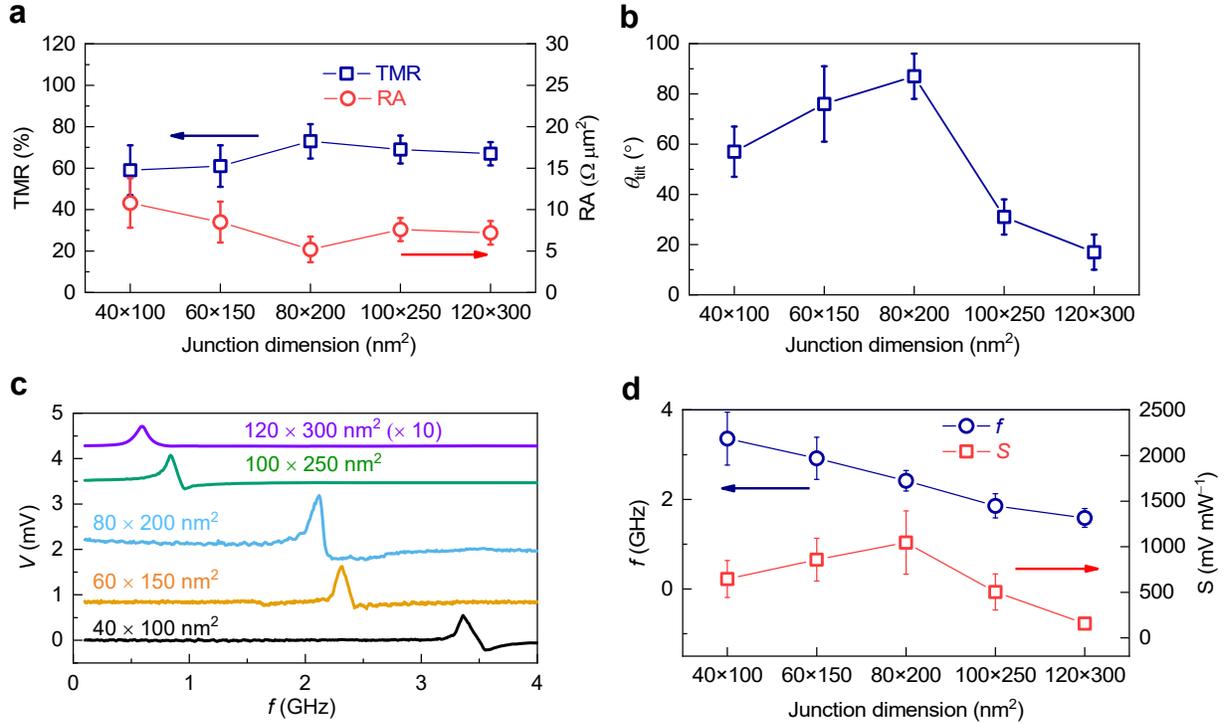

**Fig. S3| Optimization of SR dimentions varying from 40 × 100 nm² to 140 × 350 nm².** **a**, TMR and the resistance-area (RA) product versus the junction dimension. **b**, Canted equilibrium angle ($\theta_{tilt}$) calculated from Eq. (S1). The 120 × 300 nm² SR shows a canted equilibrium angle close to in-plane anisotropy. **c**, Zero-bias rectification signal voltage (*V*) at a rf power of $P_{rf}$ = −30 dBm. **d**, Frequency (*f*) of peak rectification voltage ($V_r$) and corresponding zero bias rf senstivity ($S = V_r/P$) at $P_{rf}$ = −30 dBm. Here, $P = P_{rf} \times (1-|S_{11}|^2)$, is the input power including the losses from the VNA-based $S_{11}$ measurement. The error bars in a, b, and d are the standard deviation after averaging for 5 SR devices.

**Supplementary Note 4. Contribution of the voltage controlled magnetic anisotropy**

The voltage controlled magnetic anisotropy (VCMA) is known to enhance the rf senstivity in MTJs when the electric field is applied across the device, as it excites the voltage induced ferromagnetic resonance leading to an extra antisymmetric term in the rectification spectra[3,4]. The amplitude of VCMA can be calculated as the variation of the effective magnetic anisotropy energy ($E_P$) per unit volume of the free layer from the area under the hysteresis loop $M(H)$[3,4]

$$E_P = \int_0^{M_x} H_x(M_x)\, dM_x = \frac{1}{2} M_s H_P \tag{S2}$$

where $H_P$ is the effective perpendicular anisotropy field. Figure S4a shows the change in $H_P$ with the dc bias, calculated by the shift in the conductance vs. magnetic field hysteresis loops with dc voltages. Figure S4b shows the extracted VCMA coefficient from the slope of Fig. S4a using Eq. (S2). The variation of the VCMA coefficient observed in SR-devices with the junction size is attributed to the trade-off between the effective magnetic anisotropy, the energy barrier between parallel and anti-parallel states, and the thermal stability of the MTJs[5,6], which also influences TMR and perpendicular magnetic anisotropy (PMA). The MTJs with a size of 80 × 200 nm² demonstrated the best optimization of these parameters, resulting in the best combination of VCMA, PMA, and TMR. Our systematic analysis demonstrates that optimizing the junction size, as well as the thickness of the free and MgO layers, is required to achieve canted anisotropy with a high VCMA coefficient and TMR.

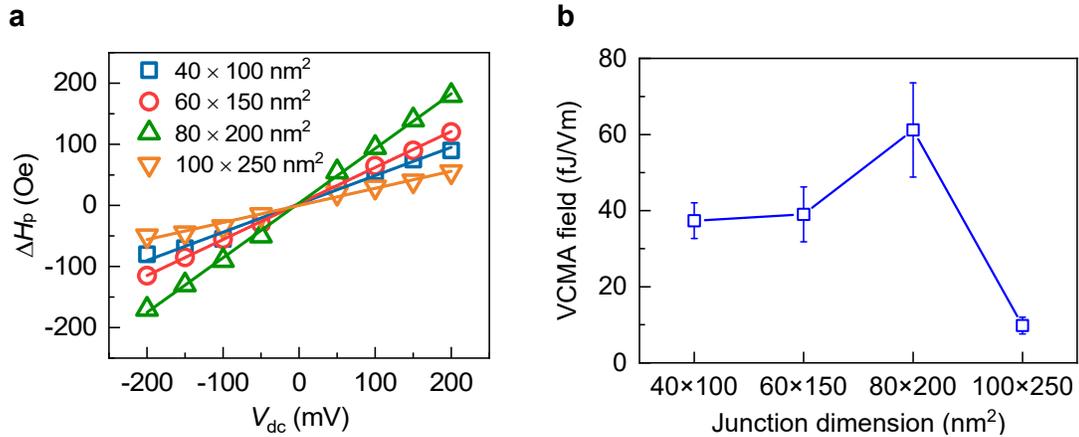

**Fig. S4| Voltage controlled magnetic anisotropy. a**, Shift of perpendicular anisotropic field ($\Delta H_P$) with applied dc bias voltages ($V_{dc}$) determined from the conductance vs. magnetic field plots. **b**, Extracted voltage-controlled magnetic anisotropy (VCMA) coefficient from the slope of Fig. S4a using Eq. (S2). The error bar shows the standard deviation after averaging for 5 SR devices.

**Supplementary Note 5. Rectified voltage vs. rf power**

Figure S5 shows the rectified voltage ($V_r$) generated by the 40 × 100 nm² and 80 × 200 nm² SRs. The corresponding senstivity is shown in Fig. 1e of the main text. The $V_r$ of 80 × 200 nm² SR saturates at a high $P_{rf}$ because of the achievement of large amplitude magnetization dynamics. In this regime, additional power is transferred to high order modes and then the sensitivity decrerases[7] as shown in Fig. 1e. The saturation voltage of the 40 × 100 nm² SR is significantly less because of the lower senstivity compared to 80 × 200 nm² SR and transition of resonant to broadband rectification response as explained in the main text.

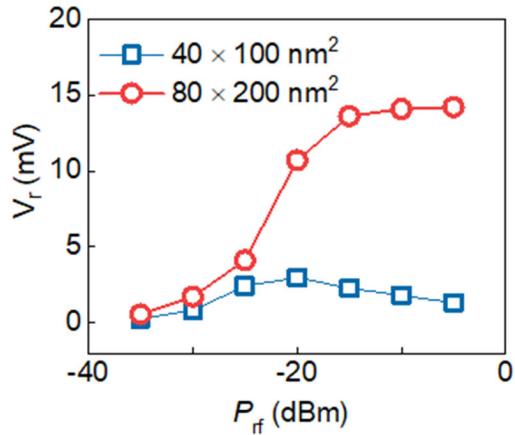

**Fig. S5| Rectified voltage with varying rf power.** The maximum rectified voltage ($V_r$) as a function of the input rf power ($P_{rf}$). The voltage saturates at $P_{rf} > -15$ dBm for the 80 × 200 nm² SR, whereas slightly decreases at $P_{rf} > -20$ dBm for the 40 × 100 nm² SR.

**Supplementary Note 6. Calibration of the rf power in the probe station measurement**

In the probe station measurement of the rectification, the impedance of the magnetic tunnel junction-based spin rectifiers is mismatched with the 50 Ω cables and connectors of the probe station, resulting in a considerable loss of rf power delivered into the SR devices. In order to calibrate the actual input power and the losses due to an impedance mismatch at the device interface, the rf power loss can be calculated using the $S_{11}$ measurement of the vector network analyser (VNA). Here, the $S_{11}$ parameter is measured in the VNA in logarithmic scale (dB) as a function of frequency with an applied power of 0 dBm, which is then converted to reflection coefficient ($|S_{11}|$) on a linear scale using the formula ($S_{11,dB} = 20 \times \log_{10}(|S_{11}|)$). Here, the reflection coefficient represents the ratio of the amplitude (or voltage) of the reflected wave to the amplitude of the incident wave.[8] The rf loss can be calculated using $|S_{11}|^2 \times 100\%$. An example of $S_{11}$ measurement is shown in Fig. S6a, where the $|S_{11}|$ is measured as 0.616 ($S_{11,dB} = -4.10$ dB on the log scale) at 2.45 GHz when a power of 0 dBm is applied from the signal generator, which is then converted to a rf power loss of $|S_{11}|^2 \times 100\%$ i.e., 38% at 2.45 GHz (Fig. S6b). In general, we measured many SR devices and found that the rf loss could vary from device to device in the range of 35-60%. Furthermore, the actual power received by the SRs can be calculated as $P = P_{rf} \times (1 - |S_{11}|^2)$, where $P_{rf}$ is the applied rf power.

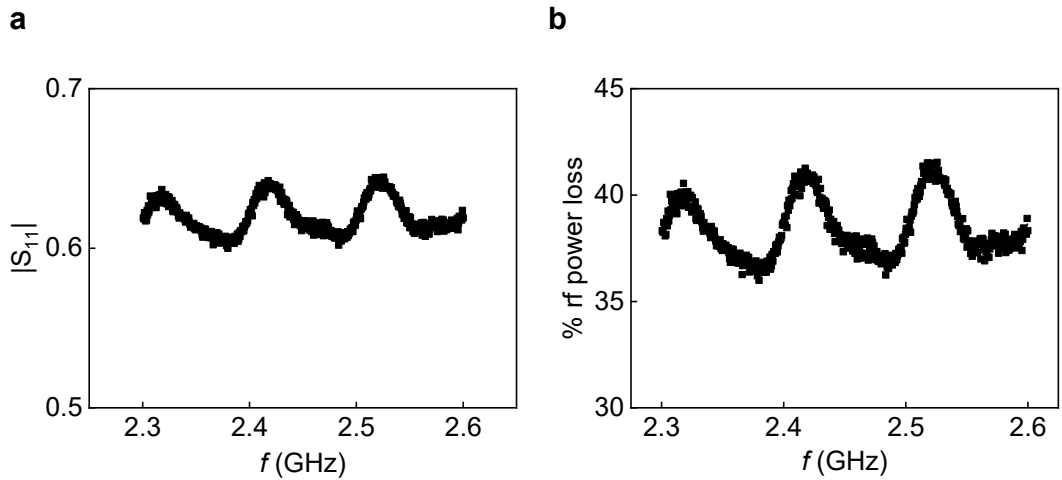

**Fig. S6| Power calibration for the probe station based measurement. a,** The linear scale magnitude of the reflection coefficient ($S_{11}$) measured in the VNA by scanning the frequency (*f*). **b,** The calculated rf power loss using the formula $|S_{11}|^2 \times 100\%$.

**Supplementary Note 7. Matched antenna design and coupling of $R_x$ and $T_x$ antenna**

The impedance of the on-chip SRs (80 × 100 nm$^2$) is measured using a vector network analyzer (VNA) in Fig. S7a. The drastic change in the impedance around 2.45 GHz is attributed to the involvement of connectors attached to the chip. However, the impedance evaluated before and after designing the matched antenna shows a robust value. We believe that the impedance matching could be improved by designing an on-chip impedance matched antenna. The performance of the matched antenna is evaluated and illustrated in Fig. S7b. The antenna can cover the entire 2.4–2.5 GHz band with the reflection coefficient ($S_{11}$) less than −10 dB as shown in Fig. S7b. The antenna is coupled with a high gain of 7.2 dBi in the far-field region, as shown by the radiation 3D pattern showing the gain in the vertical direction in Fig. S7c. For the near-field measurement (from 2.5 cm to 10 cm), the coupling between the receiving and transmitting antennas is directly estimated by the VNA measurement. Fig. S7d shows the received power measured in the VNA and corresponding peak power at 2.45 GHz is plotted in Fig. S7e. The received power at 2.45 GHz is very close to the power applied from the signal generator ($P_{rf}$) in the near field region of 2.5 cm.

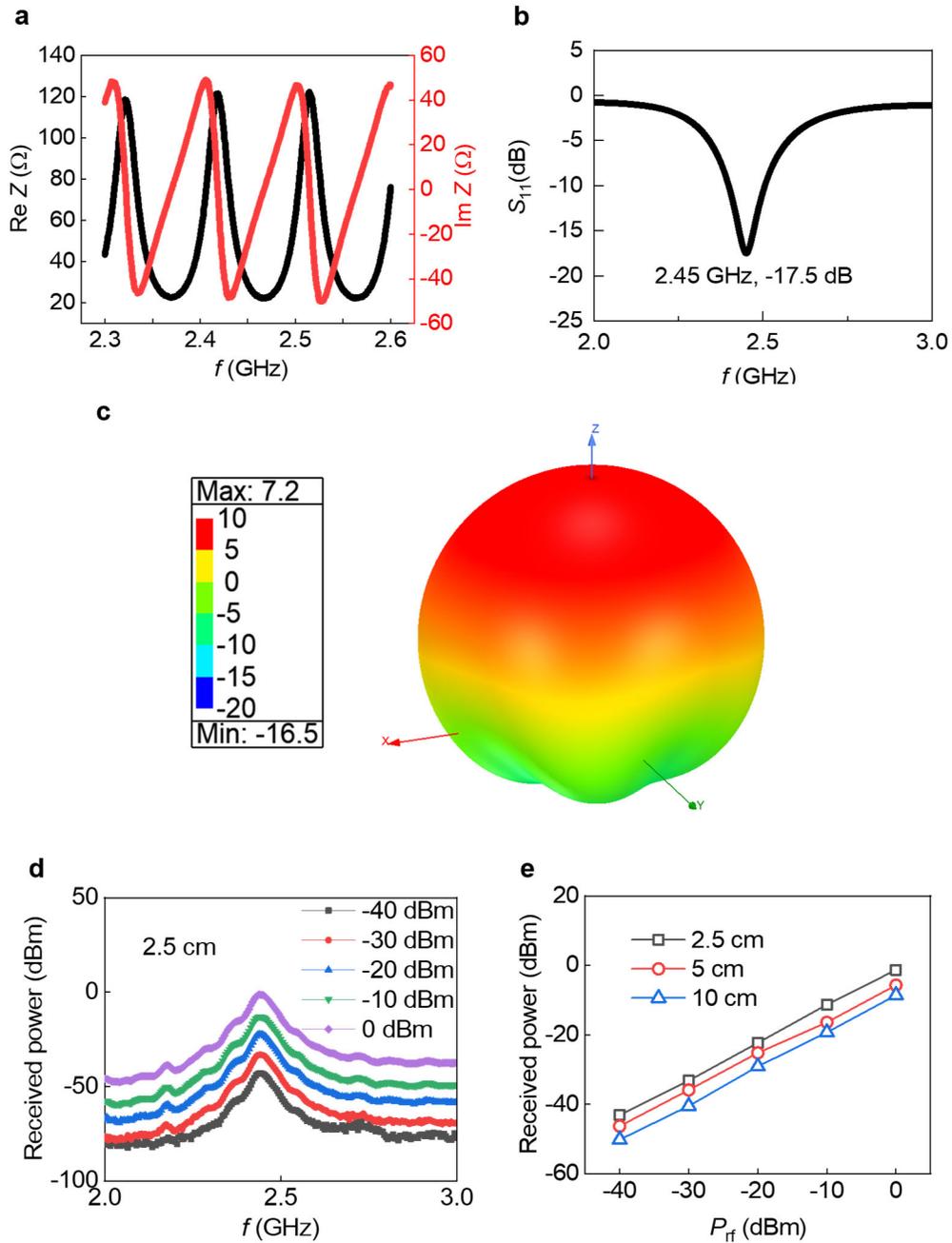

**Fig. S7| Antenna design and properties. a**, Real part of the measured impedance (Re Z) of the on-chip SR from the reflection coefficient ($S_{11}$) measured by the VNA as a function of frequency (*f*). **b**, Reflection coefficient of designed antenna showing high return loss at 2.45 GHz. **c**, 3D radiation pattern of the designed antenna using a computer simulation technology (CST) software. The color bar shows the gain on the dBi scale. **d**, The received power, measured in the VNA as a function of input rf power at 2.5 cm, shows the coupling between transmitting ($T_x$) and receiving ($R_x$) antenna. **e**, The received power at 2.45 GHz as a function of input power ($P_{rf}$).

**Supplementary Note 8. A single SR for the bandwidth up to 6 GHz**

The SR rf bandwidth is tunable with the magnetic field. By applying a magnetic field of 98 mT, a large bandwidth of 0.1–6 GHz using a single 40 × 100 nm$^2$ SR is observed as shown in Fig. S8a. Figure S8b shows the peak resonant voltage with respect to the external magnetic field applied in-plane along the major axis of the elliptical SRs, which shows that the device works in a wide range of magnetic field. However, the rectification voltage response starts to degrade at > 120 mT, because of a low TMR as the free layer becomes magnetically harder at high magnetic field (Fig. S2).

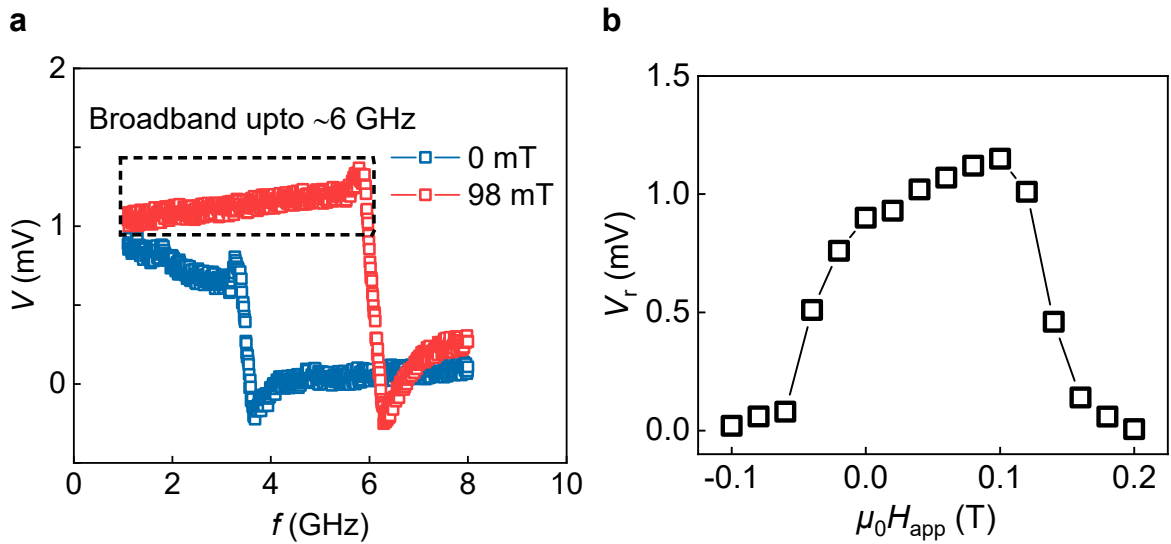

**Fig. S8| Rectification response of 40 × 100 nm$^2$ with respect to external magnetic field. a**, Broadband rectification response (shown by the dashed rectangular box) of rectified voltage (V) versus frequency (*f*) in single SR up to ~3 GHz at zero magnetic field and up to ~6 GHz using a magnetic field of 98 mT and $P_{rf}$ = −20 dBm in a 40 × 100 nm$^2$ SR. Here, the magnetic field is applied in-plane along the major axis of the elliptical SR. **b**, Variation of the peak rectified voltage ($V_r$) at $P_{rf}$ = −20 dBm with varying the in-plane magnetic field strength ($\mu_0 H_{app}$) along the major axis of the elliptical SR.

**Supplementary Note 9. Scaling of output voltage versus number of SRs connected in series**

As the number of spin rectifiers increases, the peak response becomes saturated (Fig. S9a). However, the rectification bandwidth (BW) increases as the number of SRs increases (Fig. S9b). The rectification BW is defined when the peak rectification voltage drops to half.

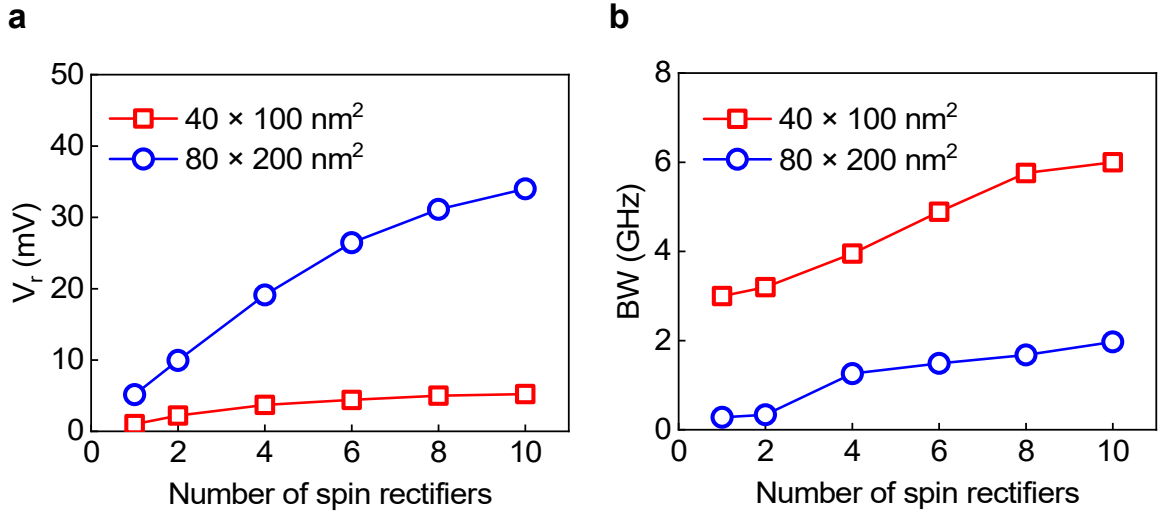

**Fig. S9| Scaling of SRs in series. a**, Peak rectified voltage ($V_r$) versus the number of spin rectifiers connected in series $P_{rf} = -20$ dBm. **b**, Rectification bandwidth (BW) at which the peak rectification voltage drops to half.

**Supplementary Note 10. Effect of VCMA and perpendicular uniaxial anisotropy constant**

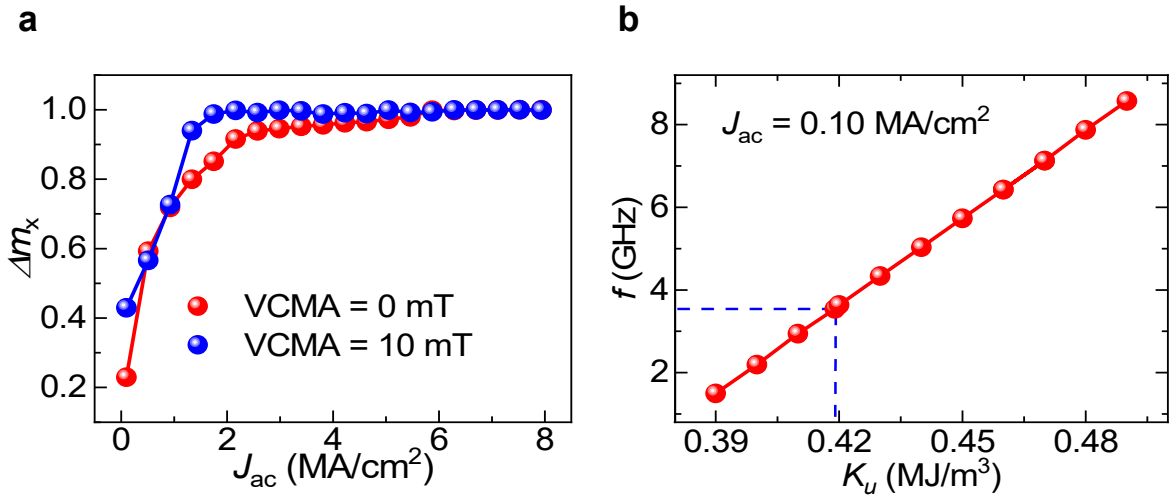

**Fig. S10| Simulation results for a 40 x100 nm² single SR. a,** Micromagnetic simulations of the VCMA driven enhancement of magnetization precession amplitude of the *x*-component ($\Delta m_x$) as a function of the rf current density ($J_{ac}$) using the same parameter of Fig. 3c in the main text. **b,** Resonance frequency (*f*) at $J_{ac}$ = 0.1 MA/cm² as a function of the perpendicular uniaxial anisotropy constant $K_u$. The dashed blue lines highlight $f \sim 3.5$ GHz, corresponding to $K_u$ = 0.419 MJ/m³, which is used in Fig. 3e for matching the experimental data. The lines are the guide for the eye.

**Supplementary Note 11. Stability of spin-rectifier arrays for long-time operation**

For testing the stability of SRs and energy harvesting module (EHM), we have stored the voltage from the SR and EHM module in the nanovoltmeter for one hour. When the source of wireless signal is available consistently in the ambient contion at $P_{rf} > -20$ dBm, the SR as well as EHM show a very consistent voltage as shown in Fig. S11. This shows that there is no charging-discharging, desynchronization of coupled SRs in series and microwave heating effect that affects the SRs performance over time.

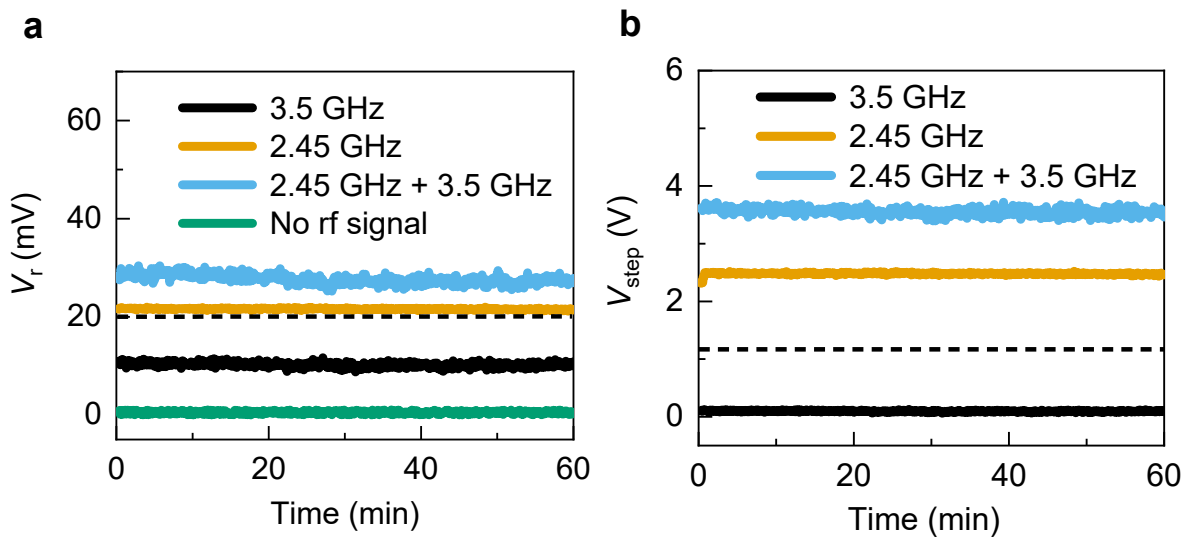

**Fig. S11| Stability test with turning the wireless signal on countinuouly for one hour. a**, SR rectification voltage ($V_r$) at $P_{rf} = -25$ dBm. **b**, EHM voltage ($V_{step}$) at $P_{rf} = -25$ dBm. The dashed lines show the minimum $V_r$ and $V_{step}$ required to turn on the dc to dc booster and 1.6 V light-emitting diode, respectively.

**Supplementary Note 12. Signal-to-noise-ratio and noise equivalent power**

The signal-to-noise ratio (SNR) determines the performance of the rf rectifiers in the noisy environment (Fig. S12a). The SNR is calculated using the formula, SNR = $P_{signal}/P_{noise}$, where the $P_{signal} = V_{dc,signal}^2/R_{dc}$ and $P_{noise} = V_{dc,noise}^2/R_{dc}$. Here, The $V_{dc,signal}$ and $V_{dc,noise}$ are the dc voltage recorded in the nanovoltmeter when the 2.45 GHz wireless signal is ON and OFF, respectively. $R_{dc}$ is the dc resistance of the devices. Moreover, the noise equivalent power (NEP) of a single SR is estimated, which characterises the noise level in quadratic detectors. The NEP is defined as the ratio of noise voltage and sensitivity. For the measurement of noise voltage ($V_{noise}$) in the spectrum analyser, we have followed the measurement setup of diode mixing noise in ref 3. We have chosen one of the SRs with a sensitivity > 1000 mV/mW. The NEP is plotted as a function of the rf power in Fig. S12b. The NEP of ~2 × $10^{-12}$ W $Hz^{0.5}$ (Fig. S12b) at $P_{rf}$ = 0.1 µW (-40 dBm) and $I_{dc}$ = 0 mA is very close to ~8 × $10^{-12}$ W $Hz^{0.5}$ or 3.6 × $10^{-12}$ W $Hz^{0.5}$ as measured in the SRs with a dc bias[3]. In general, a low NEP is observed in the SRs because of the high sensitivity, which leads to a good SNR even in SR-array configurations.

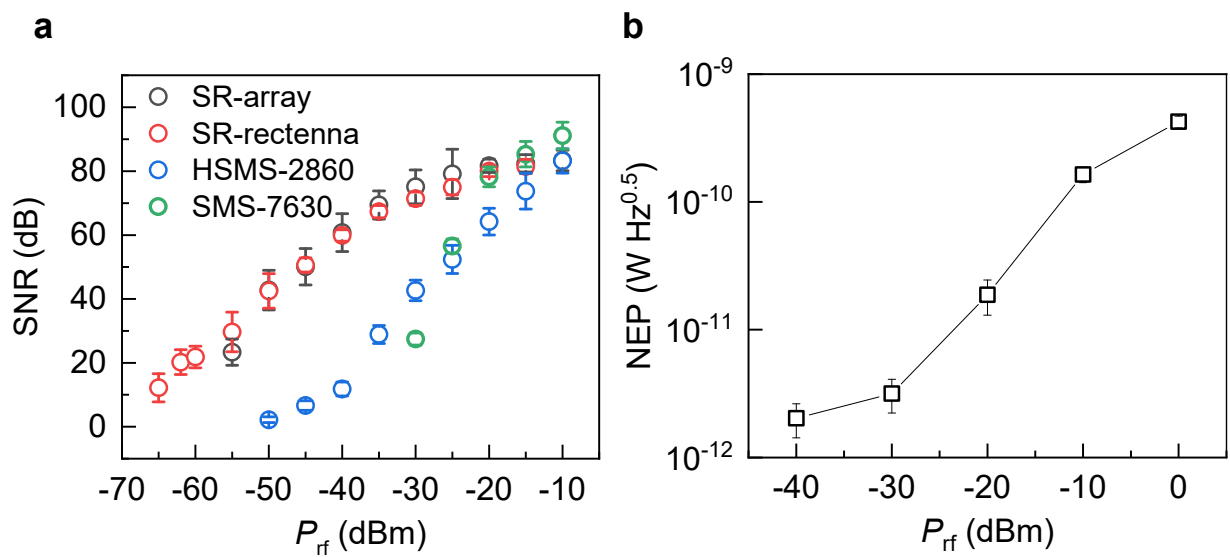

**Fig. S12| Noise perfomance of the SR-technology. a**, Signal-to-noise ratio (SNR) with varying the rf power ($P_{rf}$). The error bar defines the fluctuation of the dc voltages ($V_{dc,signal}$ and $V_{dc,noise}$) over multiple cycles, when the 2.45 GHz wireless signal is ON and OFF, respectively. **b**, Noise equivalent power (NEP) versus varying rf power. The error bar defines the standard deviation in the value of noise voltage at low frequencies.

**Supplementary Note 13. Comparison SR with the other rectifiers**

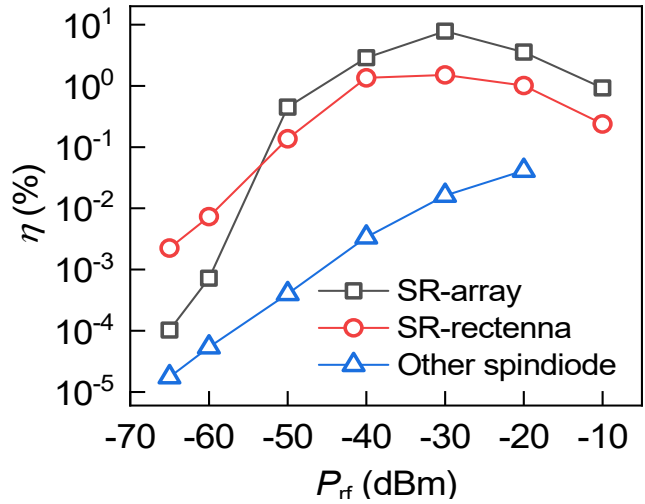

**Fig. S13| Comparison of our SR (2.45 GHz) with other SR (1.2 GHz) for $P_{rf} < -20$ dBm.** The observed efficiency ($\eta$) as a function of rf power ($P_{rf}$) is 2–3 orders higher from the previous reported efficiencies for the other spin-rectifiers with a combination of resonant[3] and broadband[7] behaviour. The lines are the guide for the eye.

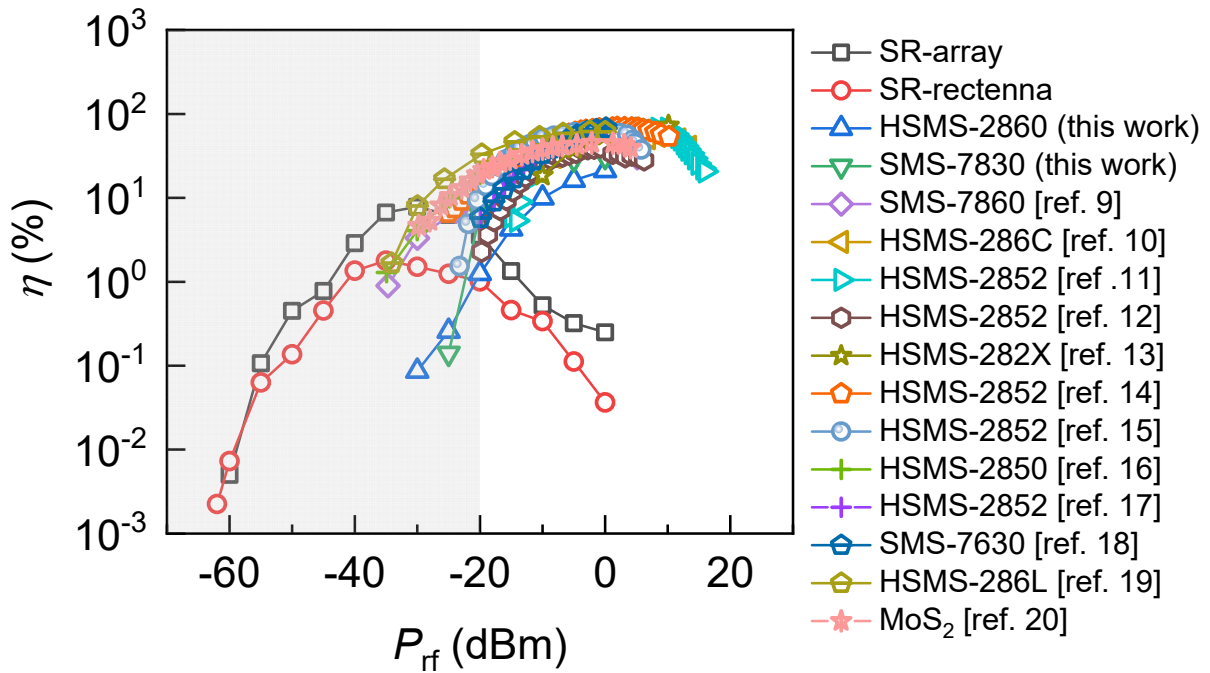

**Fig. S14| Efficiency comparison.** Comparison of efficiency reported in various highly-efficient and low-powered 2.45 GHz state-of-the-art rectifiers[9-20] (without and with taking the antenna efficiency effect).

**Supplementary Note 14. SR device and EHM setup details**

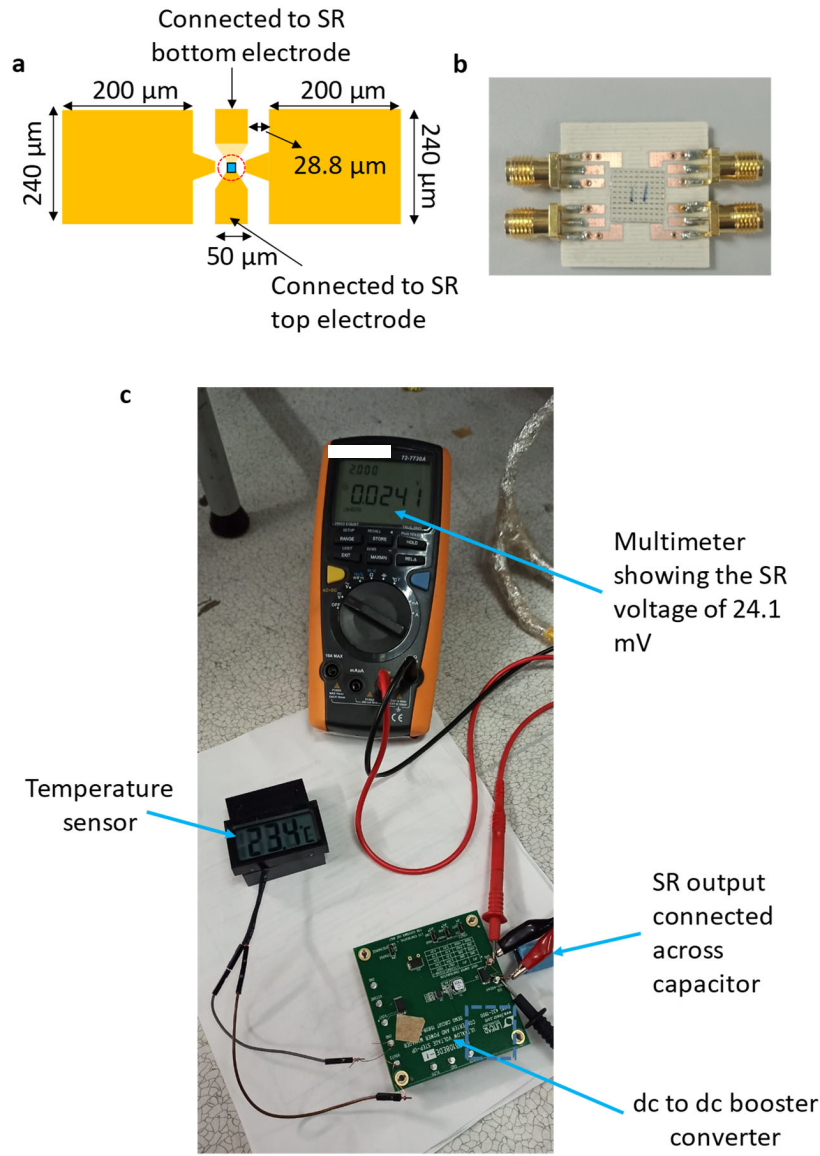

**Fig. S15: Device and experimental setup details. a**, Co-planar waveduide ground-signal-ground design over the SR devices (circled by a dashed red line) for the rf coupling. **b**, Chip, where the SR-array is connected to the rf connector, which provides a rf or dc output. The SR-devices are connected by the wirebonds. **c**, Output of the chip shown in **b** is connected across the capacitor (through black and red leads), which is connected to the input of dc to dc boost converter. A multimeter is connected to check the output of the SR-array (24.1 mV). The amplified voltage output (~3.9 V correponding to the input of 24.1 mV) from the boost converter charges the sensor, which shows a temperature of 23.4 °C.

**References**

1. Slonczewski, J. C. Currents, torques, and polarization factors in magnetic tunnel junctions. *Phys. Rev. B* **71**, 024411 (2005).
2. Zeng, Z. *et al.* Ultralow-current-density and bias-field-free spin-transfer nano-oscillator. *Sci. Rep.* **3**, 1426 (2013).
3. Fang, B. *et al.* Giant spin-torque diode sensitivity in the absence of bias magnetic field. *Nat. Commun.* **7**, 11259 (2016).
4. Zhu, J. *et al.* Voltage-induced ferromagnetic resonance in magnetic tunnel junctions. *Phys. Rev. Lett.* **108**, 197203 (2012).
5. Miriyala, V. P. K., Fong, X. & Liang, G. Influence of size and shape on the performance of VCMA-based MTJs. *IEEE Trans. Electron Devices* **66**, 944-949 (2019).
6. Shao, Y. & Khalili Amiri, P. Progress and Application Perspectives of Voltage-Controlled Magnetic Tunnel Junctions. *Adv. Mater. Technol.* **8**, 2300676 (2023).
7. Fang, B. *et al.* Experimental demonstration of spintronic broadband microwave detectors and their capability for powering nanodevices. *Phys. Rev. Appl.* **11**, 014022 (2019).
8. Caspers, F. RF engineering basic concepts: S-parameters. *arXiv preprint arXiv:1201.2346* (2012).
9. Adami, S.-E. *et al.* A flexible 2.45-GHz power harvesting wristband with net system output from− 24.3 dBm of RF power. *IEEE Trans. Microw. Theory Tech.* **66**, 380-395 (2017).
10. Chen, Y.-S. & Chiu, C.-W. Maximum achievable power conversion efficiency obtained through an optimized rectenna structure for RF energy harvesting. *IEEE Trans. Antennas Propag.* **65**, 2305-2317 (2017).
11. Kim, J. & Jeong, J. Design of high efficiency rectifier at 2.45 GHz using parasitic canceling circuit. *Microw. Opt. Technol. Lett.* **55**, 608-611 (2013).
12. Koohestani, M., Tissier, J. & Latrach, M. A miniaturized printed rectenna for wireless RF energy harvesting around 2.45 GHz. *AEU - Int. J. Electron.* **127**, 153478 (2020).
13. Mbombolo, S. E. F. & Park, C. W. in *2011 IEEE MTT-S International Microwave Workshop Series on Innovative Wireless Power Transmission: Technologies, Systems, and Applications.*  23-26 (IEEE).
14. Olgun, U., Chen, C.-C. & Volakis, J. L. Wireless power harvesting with planar rectennas for 2.45 GHz RFIDs. *2010 URSI International symposium on electromagnetic theory*, 329-331 (2010).
15. Olgun, U., Chen, C.-C. & Volakis, J. L. Investigation of rectenna array configurations for enhanced RF power harvesting. *IEEE Antennas Wirel. Propag. Lett.* **10**, 262-265 (2011).
16. Shen, S., Zhang, Y., Chiu, C.-Y. & Murch, R. A triple-band high-gain multibeam ambient RF energy harvesting system utilizing hybrid combining. *IEEE Trans. Ind. Electron.* **67**, 9215-9226 (2019).
17. Shi, Y. *et al.* An efficient fractal rectenna for RF energy harvest at 2.45 GHz ISM band. *Int. J. RF Microw.* **28**, e21424 (2018).
18. Song, C. *et al.* A high-efficiency broadband rectenna for ambient wireless energy harvesting. *IEEE Trans. Antennas Propag.* **63**, 3486-3495 (2015).
19. Wang, D. & Negra, R. Design of a rectifier for 2.45 GHz wireless power transmission. *PRIME 2012; 8th Conference on Ph. D. Research in Microelectronics & Electronics*, 1-4 (2012).
20. Zhang, X. *et al.* Two-dimensional $MoS_2$-enabled flexible rectenna for Wi-Fi-band wireless energy harvesting. *Nature* **566**, 368-372 (2019).